\documentclass[lettersize,journal]{IEEEtran}

\usepackage{cite}
\usepackage{graphicx}
\usepackage{psfrag}
\usepackage{subfigure}
\usepackage{url}
\usepackage{stfloats}
\usepackage{amsfonts,amssymb,amsmath,bm,paralist,cite,ifthen,color,amssymb}
\usepackage{amsthm}
\usepackage{array}
\usepackage{mathtools}
\usepackage{calc}
\usepackage{longtable}
\usepackage{stfloats}  
\usepackage{graphics,booktabs,color,epsfig,subfigure}
\usepackage{multirow}
\usepackage{algorithmic,algorithm}
\usepackage{subfig}
\usepackage{ifplatform}
\usepackage{xcolor}
\usepackage{preview}
\usepackage{environ}
\usepackage{graphicx}
\usepackage{epstopdf}

\usepackage{cuted}

\usepackage{tabularx}

\allowdisplaybreaks

\newtheorem{theorem}{Theorem}

\newtheorem{claim}{Claim}

\newtheorem{lem}{Lemma}
\newtheorem{rem}{Remark}

\newcommand{\beq}{\begin{equation}}
\newcommand{\eeq}{\end{equation}}

\definecolor{bluegreen}{rgb}{0.0, 0.87, 0.87}

\begin{document}

\title{A Spatial Clustering and Resource Allocation Approach for Critical Services in Massive  MIMO-Enabled Cellular Networks}
\author{Eva~Al~Hajj~Sleiman, Youssef~Hussein,~Vincent~Angilella,~Mohamad~Assaad, Senior Member, IEEE, and~Hikmet~Sari, Fellow, IEEE  

\thanks{The first two authors contributed equally to this work. 

E. Al Hajj Sleiman, Y. Hussein, and V. Angilella were with Université Paris-Saclay, CNRS, CentraleSupélec, Laboratoire des signaux et systèmes, Gif-sur-Yvette, France. 

M. Assaad is with Université Paris-Saclay, CNRS, CentraleSupélec, Laboratoire des signaux et systèmes, Gif-sur-Yvette, France. 

Hikmet Sari is with Nanjing University of Posts and Telecommunications (NJUPT), Nanjing, China.}

}



\maketitle

\begin{abstract}

	In this work, we address the challenge of providing quality-of-service (QoS) guarantees for critical real-time services in cellular networks where both real-time and non-real-time applications must be supported. Leveraging the spatial multiplexing capabilities of Massive Multiple-Input Multiple-Output (MIMO) technology, we propose a joint clustering and resource allocation framework that limits network interference and enhances the ability to meet the QoS requirements of real-time services while still serving other applications. Our approach begins by clustering active users according to the direction of their channel vectors. Specifically, a downlink probing strategy is employed to group users whose channels lie in the same direction. Channel estimation and the derivation of the corresponding signal-to-interference-plus-noise ratio (SINR) are then performed based on this clustering. Subsequently, we formulate a resource allocation optimization problem across the user groups to satisfy the required QoS for critical real-time services while accommodating the remaining users. We rigorously prove that, in general, finding the global optimum of the resource allocation problem is NP-hard, and for a special case of the formulated problem, we establish a polynomial-time solvable algorithm based on linear programming. We then propose a suboptimal solution for the general problem using a linear relaxation technique. Simulation results are provided to confirm the effectiveness of our proposed approaches.
\end{abstract}

\IEEEpeerreviewmaketitle

\section{Introduction}
\IEEEPARstart{T}{he} ever-growing demand for higher data rates, coupled with the proliferation of smart devices, has driven researchers in academia and industry to define standards capable of supporting substantial increases in data traffic and user rate requirements. For Sixth-Generation (6G) wireless communication systems, addressing traffic volume is not enough; the challenge of connecting devices under strict very low-latency (even real-time) and high-reliability constraints must also be tackled. This is particularly critical for services with real-time constraints, where meeting  quality-of-service (QoS) requirements is essential for proper operation \cite{benkhalifa}. Massive Multiple-Input Multiple-Output (MIMO) \cite{marzettanoncooperative} offers an additional avenue for improving network performance. By equipping base stations (BSs) with very large antenna arrays, mMIMO systems enable user multiplexing on shared time-frequency resources through the exploitation of spatial degrees of freedom.

In mMIMO systems, spatial degrees of freedom (DoFs) enhance reliability for critical services \cite{Sabin-1905-06205, Poposki_2018}. However, channel estimation in TDD mode relies on uplink training and reciprocity \cite{marzettatraining}, but  the training dimension is restricted by the limited coherence interval, and the low-latency constraints. Given the relation between the number of orthogonal pilots and their dimension, pilot reuse is hence inevitable, which causes pilot contamination \cite{jose}. Exploiting channel structure is therefore essential to mitigate this fundamental performance bottleneck. 
\subsection{Related Work}
Efficient beamforming and multiplexing improve reliability and latency. Recent work on non-orthogonal URLLC/eMBB coexistence has proposed spatial scheduling and beamforming for inter-service interference \cite{SoA1}, along with conformal prediction-based scheduling for dynamic traffic \cite{Popovski2023}. However, CSI acquisition remains critical: ultra-short estimation intervals force pilot reuse and cause pilot contamination \cite{jose}. In the context of 6G, studies confirm that pilot contamination governed by user coupling and large-scale fading remains a fundamental bottleneck. Exploiting spatial channel structure via user grouping thus remains a highly effective mitigation strategy \cite{SoA2}.

A large body of work leverages spatial multiplexing to improve mMIMO performance. For instance, the Joint Spatial Division and Multiplexing (JSDM) approach partitions users based on the similarity of their channel covariance eigenspaces \cite{namjoint}. Alternatively, the authors in \cite{salah} proposed partitioning users based on their channel direction inferred from downlink spatial training. In the context of critical service multiplexing, resource allocation for slicing coexisting eMBB and URLLC traffic in distributed massive MIMO was investigated in \cite{Liu2021}, employing penalized successive convex approximation for physical resource block assignment. More recently, Chong \textit{et al.} \cite{SoA2} developed a graph-theoretic framework for cell-free massive MIMO, optimizing frequency-band user grouping, pilot reuse, and power control under finite blocklength constraints. While these methods improve spectral efficiency and manage latency, they do not explore the joint optimization of resource allocation for critical services alongside standard traffic under strict spatial clustering constraints. \cite{SoA3} proposes a simulated annealing and SCA-based scheme for network slicing in user-centric CF-mMIMO to maximize sum rate under URLLC constraints. \cite{SoA4} introduces master AP selection, pilot allocation, cluster formation, and a configurable threshold for URLLC, but does not exploit massive MIMO spatial multiplexing for user grouping. \cite{SoA5} jointly optimizes power, user, and service allocations in an IRS-assisted hybrid mmWave/THz network for eMBB and URLLC coexistence. \cite{SoA6} jointly optimizes pilot length, pilot allocation, pilot power, and transmit power via geometric programming to maximize admitted devices. \cite{SoA7} derives closed-form spectral efficiency expressions for multiplexing URLLC and eMBB via network slicing in cell-free massive MIMO, without addressing user grouping or resource allocation. 

Contrary to \cite{SoA7} and the aforementioned studies, in this paper we investigate the coexistence of critical services with normal user traffic in a mMIMO system, from the perspective of both joint user grouping and resource optimization. The user grouping exploits the spatial channel directions of the users through a downlink probing method; while the resource allocation framework considers the modulation assignment to each user in addition to the subband allocation. This practical aspect of adaptive modulation allocation remains underexplored in the context of resource allocation in massive MIMO. While prior studies have investigated spatial grouping and critical services provisioning in isolation, our framework uniquely bridges these domains. By mapping analytical SINR bounds under pilot contamination directly into a mixed-integer optimization problem, we demonstrate how physical-layer spatial clustering can be explicitly tied to the joint allocation of frequencies and discrete modulation formats under strict reliability constraints.

\subsection{Contributions}
The key contributions of this work are listed below:

\begin{itemize}
    \item We first consider a user clustering strategy during the downlink probing stage that partitions users based on their spatial channel directions. Building on this partition of users, we derive the analytical channel estimates for each user, alongside the expression for the uplink Signal-to-Interference-plus-Noise Ratio (SINR).
 \item We then propose and design a joint resource optimization framework to allocate frequency and discrete modulation schemes to the established groups. This design explicitly bridges the physical-layer spatial clustering with discrete resource allocation, ensuring proportional fairness by maximizing the system sum-rate while strictly achieving the required critical services bit-rates and the hard SINR thresholds dictated by the chosen modulation formats.
 \item We theoretically demonstrate that the formulated mixed-integer optimization problem is strongly NP-hard in its search for a globally optimal solution. 
    \item We show through theoretical analysis that the associated primal problem can be transformed into a relaxed Linear Programming (LP) formulation. Furthermore, we identify a special case of this optimization problem that is tractable and can be solved to global optimality in polynomial time.
    \item We finally propose a suboptimal solution of the optimization framework and provide numerical results showing its efficiency. 
\end{itemize}

The remainder of this paper is organized as follows. Section \ref{sec:systemmodel} introduces the system model. Section \ref{sec:grouping} describes the user clustering approach, Section \ref{sec:ChannelEstimation} details the channel estimation method, and Section \ref{sec:spectralefficiency} derives the uplink SINR for each scheduled user. Section \ref{sec:frequencyallocation} formulates the frequency and modulation assignment optimization problem. Section \ref{sec:numericalresults} presents the numerical results, and Section \ref{sec:conclusion} concludes the paper.

\textbf{Notations:} Vectors and matrices are denoted by boldface lowercase (\textit{e.g.}, $\mathbf{x}$) and uppercase (\textit{e.g.}, $\mathbf{X}$) letters, respectively. The notations $\mathbf{X}^H$, $\text{Tr}\left\{ \mathbf{X}\right\}$, and $\| \mathbf{X} \|$ represent the Hermitian transpose, trace, and norm of a matrix. $\mathbf{x} \sim \mathcal{CN} \left(\boldsymbol{\mu}, \mathbf{\Sigma} \right)$ indicates that $\mathbf{x}$ is a complex circularly-symmetric Gaussian vector with mean $\boldsymbol{\mu}$ and covariance matrix $\mathbf{\Sigma}$.

\section{System Model and Group Based User Scheduling }

\subsection{System Model}\label{sec:systemmodel}

\begin{figure}[!t] 
\begin{center}
\includegraphics[width=3in]{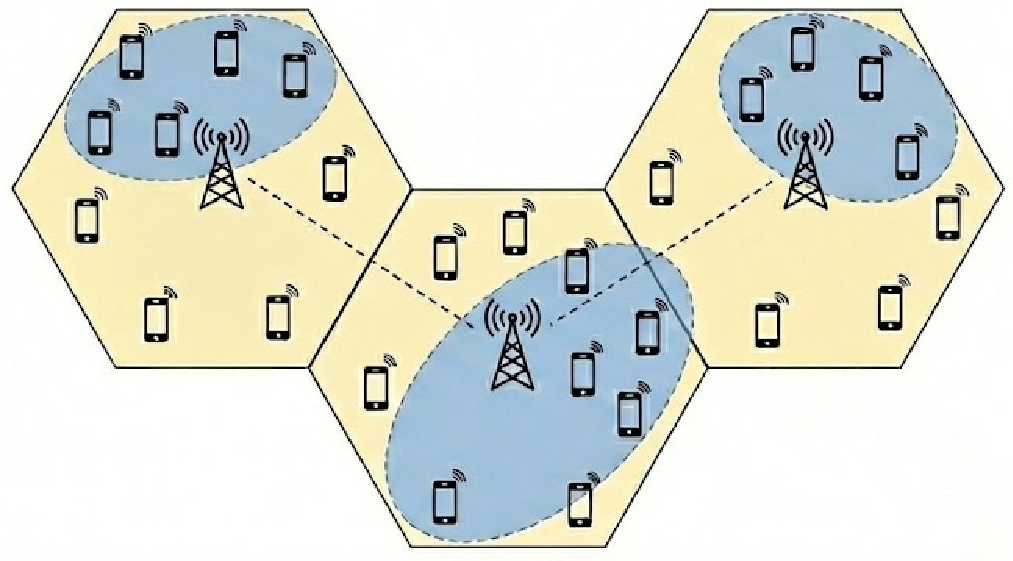}
\end{center}
\caption{System Model.} 
\label{System Model}
\end{figure}

We consider a network composed of $C$ hexagonal cells. At the center of each cell, a base station (BS) equipped with a uniform linear array (ULA) of $M$ antennas serves $K$ single omnidirectional antenna users arbitrarily distributed within the cell. We assume a mMIMO regime where $K \ll M$. Fig. \ref{System Model} depicts an example of the considered cell configuration.

The wireless channel vector from the $i$-th user in the $b$-th cell to the $j$-th BS is given by:
\begin{equation}
    \mathbf{g}_{ib}^{[j]} = \sqrt{\beta_{ib}^{[j]}} \mathbf{h}_{ib}^{[j]},
\end{equation}
where $\beta_{ib}^{[j]} \in \mathbb{R}^{+}$ denotes the large-scale fading coefficient, which accounts for path loss and shadowing. The vector $\mathbf{h}_{ib}^{[j]} \in \mathbb{C}^M\sim\ \mathcal{CN} \left( \mathbf{0},\ \mathbf{R}_{ib}^{[j]}\right)$ represents the correlated Rayleigh fading channel with covariance matrix $\mathbf{R}_{ib}^{[j]}$. Because $\mathbf{R}_{ib}^{[j]}$ is a positive semi-definite matrix,  its eigenvalue decomposition is given by $\mathbf{R}_{ib}^{[j]} = \mathbf{U}_{ib}^{[j]}  \mathbf{\Lambda}_{ib}^{[j]}  {\mathbf{U}_{ib}^{[j]}}^{H}$ where $\mathbf{U}_{ib}^{[j]} \in \mathbb{C}^{M \times n_{ib}^{[j]}}$ is a tall unitary matrix containing the eigenvectors corresponding to the $n_{ib}^{[j]}$ non-zero eigenvalues. These eigenvalues form the diagonal elements of the $n_{ib}^{[j]}\times n_{ib}^{[j]}$ diagonal matrix $\mathbf{\Lambda}_{ib}^{[j]}$. Using the Karhunen-Loève representation, the channel vector can be expressed as:
\begin{equation}
    \mathbf{h}_{ib}^{[j]} = \mathbf{U}_{ib}^{[j]} \left( \mathbf{\Lambda}_{ib}^{[j]} \right)^{\frac{1}{2}} \boldsymbol{\omega}_{ib}^{[j]},
\end{equation}
where $\boldsymbol{\omega}_{ib}^{[j]} \in \mathbb{C}^{n_{ib}^{[j]} \times 1} \sim\ \mathcal{CN} \left( \mathbf{0},\ \mathbf{I}_{n_{ib}^{[j]}} \right)$. 
We assume that the channel vectors $\{ \mathbf{h}_{ib}^{[j]} \}$ are mutually independent across all users. Furthermore, for analytical tractability, the channel covariance matrices $\{ \mathbf{R}_{ib}^{[j]} \}$ follow the one-ring local scattering model discussed in \cite{shiu}, where the $(m,p)$-th element is given by:
\begin{equation}
     \left [ \mathbf{R} \right]_{m,p} = \frac{1}{2 \Delta} \int_{- \Delta}^{\Delta}{e^{ -j\frac{2\pi}{\lambda} \left( \cos \left( \hat{\alpha} \right), \sin \left(\hat{\alpha} \right) \right)\left( \hat{\alpha} +\hat{\theta} \right) \left( \mathbf{u}_m - \mathbf{u}_p \right) }d\hat{\alpha}},
 \end{equation}
where $\Delta$ is the angular spread, $\hat{\alpha}$ is the angle of arrival (AoA), and $\lambda$ is the carrier wavelength. The vectors $\mathbf{u}_m, \, \mathbf{u}_p \in \mathbb{R}^2$ denote the spatial positions of BS antennas $m$ and $p$.

\subsection{Group Based User Scheduling}\label{sec:grouping}
We assume that the active users in the network have already been detected by their respective base stations via standard initial random access procedures. Following this initial access phase, the first step is to partition these active users into distinct groups to mitigate the effects of multi-user interference and pilot contamination. The scheme groups users whose channel vectors lie within a common spatial subspace. The first  step is to partition users into distinct groups to mitigate the effects of multi-user interference and pilot contamination. The scheme groups users whose channel vectors lie within a common spatial subspace. Similar to existing spatial-based scheduling schemes \cite{namjoint , salah}, the subspaces associated with different groups are mutually orthogonal. We assume that the subspace for each group, represented by a reference beam matrix, is known \textit{a priori} at the BS. In this paper, these reference beam matrices are generated using an $M \times M$ discrete Fourier transform (DFT) matrix $\mathbf{F}$. The elements of $\mathbf{F}$ are given by:
    \begin{equation}
        \mathbf{\left[ F \right]}_{k,l} \triangleq \frac{1}{\sqrt{M}} e^{j\frac{2 \pi}{M}k\left(l - \frac{M}{2}\right)},
    \end{equation} 
where $k, l \in \{0, \dots, M-1\}$. Each reference beam matrix $\boldsymbol{\Phi}_g\triangleq \{\phi_1,\cdots, \phi_d  \}$ is formed by selecting $d$ column vectors from the DFT matrix $\mathbf{F}$. Consequently, the total number of groups is given by $G = \lfloor \frac{M}{d} \rfloor$. By construction, any two distinct reference beam matrices are mutually orthogonal and non-overlapping, such that $\boldsymbol{\Phi}_g \cap \boldsymbol{\Phi}_{g'}=\emptyset, \, \forall g \neq g'$.

\textit{Downlink Probing:} To form the users' groups, we resort to the idea of downlink probing \cite{salah,lee}. Specifically, during the downlink probing stage, each BS transmits orthogonal reference signals (e.g., analogous to CSI-RS in 5G NR) precoded with the predefined reference beam matrices $\boldsymbol{\Phi} \triangleq \left\{\boldsymbol{\Phi}_1,\cdots, \boldsymbol{\Phi}_G  \right\}$. By measuring the received signal strength corresponding to each beam, each active user estimates the effective channel gain associated with every candidate group. The user then identifies the specific reference beam matrix $\boldsymbol{\Phi}_g$ ($g \in \{1, \dots, G\}$) that captures the majority of its channel power. Let us consider the $l$-th user in cell $r$ as the user of interest. This user determines its designated group index $g$ by verifying that the normalized power projected onto all other groups falls below a predefined threshold $\alpha \in [0, 1]$:
\begin{equation}\label{eq:inegalpha}
    \frac{\left\| \boldsymbol{\Phi}_{g'} \mathbf{g}_{lr}^{[r]} \right\|}{\left\| \mathbf{g}_{lr}^{[r]} \right\|} \leq \alpha, \quad \forall g' \in \{1, \dots, G\} \setminus \{g\}.
\end{equation}
Following this downlink evaluation, if a user satisfies the condition in \eqref{eq:inegalpha} for a unique group $g$, it feeds back the index $g$ to its serving BS via an uplink control channel. Users whose channel power is too dispersed to satisfy this strict spatial isolation criterion for any group are deferred and not scheduled during the current coherence interval. In a practical dynamic system, these deferred users are reconsidered in subsequent coherence intervals as their fast-fading channel realizations evolve, or they are served via standard orthogonal fallback resources (e.g., dedicated time slots) by the higher-layer MAC scheduler to prevent long-term starvation.

It is important to emphasize that group assignment relies on this strict spatial isolation threshold rather than a simple maximum-power association rule. If a highly dispersed user were simply assigned to the group containing their maximum signal power, they would still leak significant interference into the subspaces of other groups. Because pilot sequences are reused, this leakage would severely degrade channel estimation and inflate intra-cell interference, jeopardizing the strict reliability guarantees required by critical services. The threshold $\alpha$ ensures that inter-group spatial leakage remains strictly bounded, protecting the integrity of the orthogonal subspaces.

\begin{rem}
It is worth noting that the signaling overhead associated with this downlink probing and uplink feedback procedure is relatively low. The BS is required to transmit $G = \lfloor M/d \rfloor$ orthogonal reference signals. While this scales with the number of antennas $M$, the cluster size $d$ effectively reduces the required probing dimension by a factor of $d$ compared to a full $M$-dimensional spatial sweep. Furthermore, to report the preferred group index $g$, each user only needs to transmit $\lceil \log_2(G) \rceil$ bits on the uplink control channel (e.g., 2 bits for $G=4$), making the feedback highly scalable.
\end{rem}

\subsection{Pilot Transmission and Channel Estimation}\label{sec:ChannelEstimation}
Once the groups have been identified using the aforementioned downlink probing strategy and $\tau$ users have been scheduled per group, we incorporate the group index into the channel vector of each scheduled user for the purpose of channel estimation. To this end, the channel vector from the $l$-th user in group $k$ of cell $b$ to the $r$-th BS is rewritten as:
\begin{equation}
    {\mathbf{g}}_{lb}^{\left[rk \right]} = \sqrt{\beta_{lb}^{\left[ rk \right]}}\mathbf{h}_{lb}^{\left[rk \right]},
\end{equation}
where $
  \mathbf{h}_{lb}^{\left[rk \right]}  =  \mathbf{U}_{lb}^{\left[rk \right]} \left( \mathbf{\Lambda}_{lb}^{\left[rk \right]} \right) ^{\frac{1}{2}} \boldsymbol{\omega}_{lb}^{\left[ rk \right]}, \,\, \boldsymbol{\omega}_{lb}^{\left[ rk \right]}\sim\ \mathcal{CN}(\mathbf{0},\ \mathbf{I}_{n_{lb}^{[rk]}}) $. In other words, $\mathbf{g}_{lb}^{\left[ rk \right]} \sim\ \mathcal{CN}\left(\mathbf{0},\ \mathbf{R}_{lb}^{[rk]}\right) $ with $     \mathbf{R}_{lb}^{[rk]} = \beta_{lb}^{[rk]} \mathbf{U}_{lb}^{[rk]}\mathbf{\Lambda}_{lb}^{[rk]}\mathbf{U}_{lb}^{[rk]^\dagger}$.

Assuming uplink/downlink channel reciprocity, the downlink channel can be estimated via uplink training. Within a given group, the $\tau$ users are assigned mutually orthogonal pilot vectors. However, these $\tau$ pilot sequences are reused and allocated to users belonging to different groups within the same cell, as well as to users in neighboring cells. During the training stage, the received pilot signal at the $r$-th BS is given by:\footnote{During the training stage, for the sake of simplicity and without loss of generality, we normalize the pilot power to $1$.}

\begin{equation}
\mathbf{Y}_p^{\left[ r \right]} = \sum\limits_{b=1}^{C}\sum\limits_{k=1 }^{G}\sum\limits_{l=1}^{\tau} {\mathbf{g}_{lb}^{\left[ rk \right]} \mathbf{q}_{\pi_{lb}^k}^H} + \mathbf{N}_p,
\end{equation}
where $\pi_{lb}^k$ denotes the index of the pilot vector assigned to user $l$ in the $k$-th group of cell $b$. The term $\mathbf{N}_p \in \mathbb{C}^{M \times \tau}$ is the AWGN matrix, whose entries follow a complex circularly-symmetric Gaussian distribution, i.e., $\sim \mathcal{CN}(0,\sigma^2)$, where $\sigma^2$ is the noise variance.

The channel estimate for a given user is obtained by correlating the received training signal with its known pilot sequence. The minimum mean square error (MMSE) estimator for the direct link channel vector of the $l$-th user associated with group $k$ in the target cell $r$ is given by:
\begin{equation}
    \hat{\mathbf{g}}_{lr}^{\left[ rk \right]} = \mathbf{\Theta}_{lr}^{\left [ rk \right]} \left( \hat{\mathbf{g}}_{lr}^{\left[ rk\right]} \right)^{LS},
\end{equation}
where
\begin{equation}
     \mathbf{\Theta}_{lr}^{[rk]} =  \mathbf{R}_{lr}^{[rk]} \left( \rho\mathbf{I}_M + \sum\limits_{b=1}^{C}\sum\limits_{j =1}^G\mathbf{R}_{lb}^{[rj]} \right) ^{-1}  ,
\end{equation}
$\rho = \sigma^2 / \tau$, and $\left( \hat{\mathbf{g}}_{lr}^{\left[ rk \right]} \right)^{LS}$ is the least square (LS) estimate of the channel, defined as:
\begin{align}
     \left( \hat{\mathbf{g}}_{lr}^{\left[ rk \right]} \right)^{LS} &= \mathbf{Y}_p^{\left[ r \right]} \mathbf{q}_l = \sum\limits_{b=1}^{C}\sum\limits_{j \in \nu(b)}{\mathbf{g}_{lb}^{\left[ rj \right]}} +\mathbf{N}_p \mathbf{q}_l .
\end{align}

We define $\nu(b) \subseteq \{1, \ldots, G\}$ as the set of indices for the user groups that are active and scheduled for transmission in cell $b$. Consequently, the mean square error (MSE) of the channel estimate is given by:
\begin{align}
    {MSE}_{lr} &=\mathop{\mathbb{E}} \left[ {\left\| \mathbf{g}_{lr}^{\left[ rk \right ]} - \hat{\mathbf{g}}_{lr}^{\left[ rk \right ]} \right\|}^2 \right]\\ &= Tr \left\{ \mathop{\mathbb{E}} \left[ \mathbf{g}_{lr}^{\left[ rk \right ]}\mathbf{g}_{lr}^{\left[ rk \right ]H} - \hat{\mathbf{g}}_{lr}^{\left[ rk \right ]} \hat{\mathbf{g}}_{lr}^{\left[ rk \right ]H}  \right] \right\},
\end{align}
where
\begin{equation}
\begin{split}
\mathop{\mathbb{E}}\!\left[\hat{\mathbf{g}}_{lr}^{[rk]} \hat{\mathbf{g}}_{lr}^{[rk]H}\right]
= \mathop{\mathbb{E}}\!\Big[
\mathbf{\Theta}_{lr}^{[rk]}
\Big( \sum_{b=1}^{C} \sum_{j \in \nu(b)} \mathbf{g}_{lb}^{[rj]} + \mathbf{N}_p \mathbf{q}_l \Big) \\
\qquad\qquad\qquad \times
\Big( \sum_{b=1}^{C} \sum_{j \in \nu(b)} \mathbf{g}_{lb}^{[rj]H} + \mathbf{q}_l^{H} \mathbf{N}_p^{H} \Big)
\mathbf{\Theta}_{lr}^{[rk]H}
\Big].
\end{split}
\label{eq:valeqc}
\end{equation}
Note that we have:
\begin{equation}
\begin{split}
\mathbb{E}\!\Big[
\Big(  {}&\sum_{b=1}^{C}\sum_{j \in \nu(b)} \mathbf{g}_{lb}^{[rj]} + \mathbf{N}_p \mathbf{q}_l \Big)
\Big( \sum_{b=1}^{C}\sum_{j \in \nu(b)} \mathbf{g}_{lb}^{[rj]H} + \mathbf{q}_l^{H}\mathbf{N}_p^{H} \Big)
\Big]\\
= {}& \sum_{b=1}^{C}\sum_{j \in \nu(b)} \sum_{b'=1}^{C}\sum_{j' \in \nu(b)}
\mathbb{E}\!\left[ \mathbf{g}_{lb}^{[rj]} \mathbf{g}_{lb'}^{[rj']H} \right]
+ \mathbb{E}\!\left[ \mathbf{N}_p \mathbf{q}_l \mathbf{q}_l^{H} \mathbf{N}_p^{H} \right] \\
= {}& \rho\mathbf{I}_M
+ \sum_{b=1}^{C}\sum_{j \in \nu(b)} \sum_{b'=1}^{C}\sum_{j' \in \nu(b)}
\mathbb{E}\!\Bigg[
\sqrt{\beta_{lb}^{[rj]}\beta_{lb'}^{[rj']}}
\,\mathbf{U}_{lb}^{[rj]}\\
{}&\qquad \qquad \times 
\big(\mathbf{\Lambda}_{lb}^{[rj]}\big)^{\!1/2}\boldsymbol{\omega}_{lb}^{[rj]}\boldsymbol{\omega}_{lb'}^{[rj']H}
\big(\mathbf{\Lambda}_{lb'}^{[rj']}\big)^{\!1/2}
\mathbf{U}_{lb'}^{[rj']H}
\Bigg] \\
= {}& \rho\mathbf{I}_M
+ \sum_{b=1}^{C}\sum_{j \in \nu(b)}
\beta_{lb}^{[rj]}\,\mathbf{U}_{lb}^{[rj]}\mathbf{\Lambda}_{lb}^{[rj]}\mathbf{U}_{lb}^{[rj]H} \\
= {}&\rho \mathbf{I}_M
+ \sum_{b=1}^{C}\sum_{j \in \nu(b)} \mathbf{R}_{lb}^{[rj]}.
\end{split}
\label{eq:expec_val}
\end{equation}
By substituting \eqref{eq:expec_val} into \eqref{eq:valeqc}, it follows that: 
\begin{equation}
\begin{split}
    \mathop{\mathbb{E}} \left[\hat{\mathbf{g}}_{lr}^{\left[ rk \right ]} \hat{\mathbf{g}}_{lr}^{\left[ rk \right ]H} \right] =&  \mathbf{R}_{lr}^{[rk]} \Big(
     \mathbf{I}_M + \sum\limits_{b=1}^{C}\sum\limits_{j \in \nu \left( b \right)}\mathbf{R}_{lb}^{[rj]} \Big) ^{-1}  \mathbf{R}_{lr}^{[rk]}.
     \end{split}
\end{equation}
Hence, the $MSE_{lr}$ becomes:
\begin{equation}
    {MSE}_{lr}=  Tr \left\{ \mathbf{R}_{lr}^{[rk]} - \mathbf{C}_{lr}^{[rk]} \right\} .
\end{equation}
Consequently, the MMSE estimate of the channel vector can be explicitly expressed as: 
\begin{equation}\label{eq:fina_mmse}
\begin{aligned}
        \hat{\mathbf{g}}_{lr}^{\left[ rk \right]} ~&= \mathbf{R}_{lr}^{[rk]} \left(
     \mathbf{I}_M + \sum\limits_{b=1}^{C}\sum\limits_{j \in \nu \left( b \right)}\mathbf{R}_{lb}^{[rj]} \right) ^{-1} \mathbf{Y}_p^{\left[ r \right]} \mathbf{q}_l \\
     ~&=\mathbf{R}_{lr}^{[rk]} \left(
     \mathbf{I}_M + \sum\limits_{b=1}^{C}\sum\limits_{j \in \nu \left( b \right)}\mathbf{R}_{lb}^{[rj]} \right) ^{-1}\\& \qquad \times\left( \sum\limits_{b=1}^{C}\sum\limits_{j \in \nu \left( b \right)}{\mathbf{g}_{lb}^{\left[ rj \right]}} +\mathbf{N}_p \mathbf{q}_l \right) .
\end{aligned}
\end{equation}
In this paper, we consider imperfect channel estimation. Let $\tilde{\mathbf{g}}_{lr}^{\left[ rk \right]} \triangleq  \mathbf{g}_{lr}^{\left[ rk \right]} - \hat{\mathbf{g}}_{lr}^{\left[ rk \right]}$denote the channel estimation error. Due to the properties of MMSE estimation, $ \hat{\mathbf{g}}_{lr}^{\left[ rk \right]}$ and  $\tilde{\mathbf{g}}_{lr}^{\left[ rk \right]}$ are uncorrelated and are respectively distributed as:
\begin{equation}\label{eq:distr_channel}
\begin{aligned}
  \hat{\mathbf{g}}_{lr}^{\left[ rk \right]} ~& \sim\ \mathcal{CN}\left(\mathbf{0},\ \mathbf{C}_{lr}^{[rk]}\right),  \\
       \tilde{\mathbf{g}}_{lr}^{\left[ rk \right]} ~&\sim\ \mathcal{CN}\left(\mathbf{0},\ \mathbf{R}_{lr}^{[rk]} - \mathbf{C}_{lr}^{[rk]} \right) .
\end{aligned}
\end{equation}
It is worth noting that, although pilot sequences are reused within a given cell, users assigned the same pilot have their channel vector powers confined to mutually orthogonal subspaces. Therefore, intra-cell pilot contamination is significantly mitigated. However, the channel estimate will still suffer degradation from inter-cell pilot contamination. 

\begin{rem}
A lower value for the threshold $\alpha$ reduces intra-cell pilot contamination interference, leading to a more accurate channel estimate and, consequently, enhanced spectral efficiency. Notably, when $\alpha = 0$, intra-cell pilot contamination is completely eliminated, leaving only inter-cell pilot contamination as the source of interference.
\end{rem}

\subsection{SINR expression}
\label{sec:spectralefficiency}
Let $r$ denote the reference BS. The received uplink complex baseband signal is:
\begin{equation}\label{eq:uplinksignal}
    \mathbf{y}_u^{[r]} = \sum\limits_{b=1}^{C}\sum\limits_{k \in \nu \left( b \right)}\sum\limits_{i=1}^{\tau} {\mathbf{g}_{ib}^{[rk]}s_{ib}^{[bk]} }+\mathbf{n}_u,
\end{equation}
where $s_{ib}^{[bk]} \sim \mathcal{CN}(0, 1)$ is the transmitted symbol and $\mathbf{n}_u \sim \mathcal{CN}(\mathbf{0}, \sigma^2 \mathbf{I}_M)$ is the AWGN vector. To detect the symbol from the $l$-th user in the $k$-th group, BS $r$ applies the matched filter $\left( \mathbf{\Phi}_k \mathbf{\Phi}_k^H\hat{\mathbf{g}}_{lr}^{\left[ rk \right]} \right)^H$ to \eqref{eq:uplinksignal}, yielding:
\begin{equation}
  \hat{s}_{lr}^{[rk]}=\sum\limits_{b,\, j\in\nu\left(b \right )}\sum\limits_{i=1}^{\tau}{\hat{\mathbf{g}}_{lr}^{[rk]H}}\boldsymbol{\Phi }_k\boldsymbol{\Phi }_k^H\boldsymbol{\mathbf{g}}_{ib}^{[rj]}s_{ib}^{[bk]}  + {\hat{\mathbf{g}}_{lr}^{[rk]H}}\boldsymbol{\Phi }_k\boldsymbol{\Phi }_k^H\boldsymbol{\mathbf{n}}_u.
\end{equation}

\begin{lem}
\label{lemme:sinr}
Using the matched filter $\left( \mathbf{\Phi}_k \mathbf{\Phi}_k^H\hat{\mathbf{g}}_{lr}^{\left[ rk \right]} \right)^H$, the uplink SINR of the $l$-th user in the $k$-th group within cell $r$ is lower-bounded by:
\begin{equation}\label{SINRexpr}
\begin{aligned}
     SINR_{lr}^{\left[ k \right]} = &\frac{\left| Tr \left\{ \mathbf{C}_{lr}^{\left[ rk \right]} \mathbf{\Phi}_k \mathbf{\Phi}_k^H \right\} \right| ^2}{\substack{Tr \left\{ \mathbf{C}_{lr}^{\left[ rk \right]} \mathbf{\Phi}_k \mathbf{\Phi}_k^H\mathbf{C}_{lr}^{\left[ rk \right]} \mathbf{\Phi}_k \mathbf{\Phi}_k^H\right\}  + \sum\limits_{\substack{b=1 \\ b \neq r}}^C\sum\limits_{ j \in \nu \left( b \right)} \mathcal{I}_{\text{inter},lr}^{[k|bj]}  \\ + \sum\limits_{\substack{ j \in \nu \left( r\right) \\ j \neq k}}  \mathcal{I}_{\text{intra},lr}^{[k|j]}  + \sum\limits_{b=1}^{C} \sum\limits_{j \in \nu \left( b \right)} \sum\limits_{\substack{i=1 \\ i \neq l}}^{\tau}  \mathcal{I}_{\text{orth},lr}^{[k|bji]} \\ + \sum\limits_{b=1}^{C}\sum\limits_{j \in \nu \left( b \right)}\sum\limits_{i=1}^{\tau} \mathcal{I}_{\text{err},lr}^{[k|bji]} + \sigma^2 Tr \left\{ \mathbf{C}_{lr}^{\left[ rk \right]} \mathbf{\Phi}_k \mathbf{\Phi}_k^H \right\}}},
\end{aligned}
    \end{equation}
where 
\begin{equation}
    \begin{aligned}
        \mathcal{I}_{\text{inter},lr}^{[k|bj]}=&Tr \left\{ \mathbf{C}_{lr}^{\left[ rk \right]} \mathbf{\Phi}_{k} \mathbf{\Phi}_{k}^{H} \mathbf{C}_{lb}^{\left[ rj \right]}\mathbf{\Phi}_{k} \mathbf{\Phi}_{k}^{H} \right\} \\&\quad+ \left| Tr \left\{ \mathbf{\Theta}_{lb}^{\left[ rj \right]} \mathbf{R}_{lr}^{\left[ rk \right] }\mathbf{\Phi}_{k} \mathbf{\Phi}_{k}^{H}\right\} \right| ^{2},\\
        \mathcal{I}_{\text{intra},lr}^{[k|j]}=&\Big( Tr \left\{ \mathbf{C}_{lr}^{\left[ rk \right]} \mathbf{\Phi}_{k} \mathbf{\Phi}_{k}^{H} \mathbf{C}_{lr}^{\left[ rj \right]}\mathbf{\Phi}_{k} \mathbf{\Phi}_{k}^{H} \right\} \\&\quad+ \left| Tr \left\{ \mathbf{\Theta}_{lr}^{\left[ rj \right]} \mathbf{R}_{lr}^{\left[ rk \right] }\mathbf{\Phi}_{k} \mathbf{\Phi}_{k}^{H}\right\} \right| ^{2} \Big) \alpha^4,\\
        \mathcal{I}_{\text{orth},lr}^{[k|bji]} =& Tr \left\{ \mathbf{C}_{lr}^{\left[ rk \right]}\mathbf{\Phi}_k \mathbf{\Phi}_k^H  \mathbf{C}_{ib}^{\left[ rj \right]} \mathbf{\Phi}_k \mathbf{\Phi}_k^H \right\},\\
        \mathcal{I}_{\text{err},lr}^{[k|bji]}=& Tr \left\{ \mathbf{C}_{lr}^{\left[ rk \right] }\mathbf{\Phi}_k \mathbf{\Phi}_k^H \left( \mathbf{R}_{ib}^{\left[ rj\right]} - \mathbf{C}_{ib}^{\left[ rj \right]} \right)\mathbf{\Phi}_k \mathbf{\Phi}_k^H \right\}.
    \end{aligned}
\end{equation}
\end{lem}

\begin{proof}
The detailed proof is provided in Appendix \ref{ProofSinr}.
\end{proof}
\begin{lem}
\label{lemme:larged}

For a large number of antennas $M$, the uplink SINR is bounded by \cite{salah}:
\begin{equation}\label{sinr_assyptregime}
    SINR_{lr}^{\left[ k \right]} = \frac{\left( \beta_{lr}^{\left[ rk \right]} \right)^2}{\sum\limits_{\substack{b=1 \\ b \neq r}}^C\sum\limits_{j \in \nu \left( b \right)}{\left( \beta_{lb}^{\left[ rj \right]} \right)^2} + \sum\limits_{\substack{j \in  \nu \left( r \right) \\ j \neq k}}{\left( \beta_{lr}^{\left[ rj\right]}\right)^2 \alpha^4} }.
\end{equation}
\end{lem}
\begin{proof}
The proof follows directly from Lemma \ref{lemme:sinr}.
\end{proof}

\begin{rem}
The denominator in \eqref{sinr_assyptregime} comprises inter-cell interference (first term) and intra-cell interference (second term).
\end{rem}

\begin{rem}
As $\alpha \rightarrow 0$, intra-cell interference vanishes. However, setting $\alpha = 0$ severely restricts user admission and fundamentally alters system design.
\end{rem}

\section{Frequency and modulation allocation}

\label{sec:frequencyallocation}
In the previous section, we derived the SINR expression for the grouping-based scheme assuming a TDD system. Due to the limited channel coherence time, resources dedicated to uplink training are limited and must be reused, causing intra-cell and inter-cell pilot contamination interference. Although one can argue that intra-cell pilot contamination interference disappears in the asymptotic regime by setting $\alpha=0$, this is a rather strong assumption which does not always hold for practical scenarios. It is therefore of utmost importance to investigate approaches to mitigate the effects of intra-cell and inter-cell interference. A viable solution is to resort to joint radio resource and modulation allocation, which represents the main contribution of this paper. 

For the reader's convenience, the primary sets, indices, system parameters, and decision variables used throughout our mathematical formulation are summarized in Table \ref{tab:notation}.

\begin{table}[ht!]
\caption{Summary of Main Notations}
\label{tab:notation}
\centering
\renewcommand{\arraystretch}{1.2}
\begin{tabularx}{\columnwidth}{l X}
\hline\hline
\textbf{Symbol} & \textbf{Description} \\
\hline
$b, r \in \{1, \dots, C\}$ & Index of cells ($r$ denotes the reference cell) \\
$j, k \in \{1, \dots, G\}$ & Index of spatial user groups \\
$i, l \in \{1, \dots, K_j\}$ & Index of active users within a group \\
$f \in \{1, \dots, F\}$ & Index of available frequency resources \\
$m \in \{1, \dots, L\}$ & Index of available modulation formats \\
$\nu(b)$ & Set of active spatial groups scheduled in cell $b$ \\
$\mathcal{U}(j,b)$ & Set of critical services users in group $j$, cell $b$ \\
$M, K$ & BS antennas count and total active users per cell \\
$\alpha$ & Spatial isolation threshold for user grouping \\
$\tau$ & Maximum number of scheduled users per group \\

$\sigma^2$ & Additive white Gaussian noise variance \\

$w_{ib}^j$ & Fairness priority weight for user $i$ \\
$R_{ib}^j$ & Minimum data rate constraint for a critical user \\
$R_m, \theta_m$ & Data rate and minimum required SINR threshold for modulation $m$ \\
$x_{ib}^{jfm} \in \{0, 1\}$ & Binary allocation variable (user $i$ in cell $b$, group $j$). \\
\hline\hline
\end{tabularx}
\end{table}

\subsection{Problem Formulation}
The proposed optimization problem consists of assigning $F$ available frequencies and $L$ modulation schemes to maximize the total network throughput, subject to strict rate requirements for critical services users and SINR thresholds for the assigned modulation schemes. The mathematical formulation of the optimization problem is written as:
\begin{subequations}\label{Opt_proble_frerq_modul}
\begin{align}
 \max_{\textbf{x}} ~&  \sum\limits_{b=1}^{C}\sum\limits_{j\in\nu(b)}\sum\limits_{i=1}^{K_j}{w_{ib}^j\sum\limits_{f=f_1}^{f_F}\sum\limits_{m=m_1}^{m_L}{x_{ib}^{jfm}R_m}} \label{mip:cout} \\
\textrm{s.t.}  ~&   \sum\limits_{m=m_1}^{m_L}x_{ib}^{jfm}  \leq  1, \; \forall b,  \;  \forall j , \;  \forall i,  \;  \forall f,  \label{mip:1modulation}  \\
 ~&   \sum\limits_{i=1}^{K_j}\sum\limits_{m=m_1}^{m_L}x_{ib}^{jfm}  \leq  \tau,  \forall b , \;  \forall j, \;  \forall f,  \label{mip:tau} \\
 ~&   {SINR}_{ib}^{jf}  \geq  x_{ib}^{jfm}\theta_m, \; \forall b ,  \;  \forall j, \;  \forall i ,  \; \forall f ,\forall m , \label{mip:sinr} \\
 ~&  \sum\limits_{f=f_1}^{f_F}\sum\limits_{m=m_1}^{m_L}{x_{ib}^{jfm}R_m}  \geq  R_{ib}^j, \; \forall\ i\in\ \mathcal{U}\left(j,b\right); \; \forall b ,  \;  \forall j, \label{mip:urllc}\\
  ~&  \sum\limits_{i\in \Pi\left(j,b,p\right)}\sum\limits_{m=m_1}^{m_L}{x_{ib}^{jfm}}  \leq  1, \; \forall b ,  \;  \forall j,  \; \forall f ,  \; \forall p , \label{mip:pilotlimit}
\end{align}
\end{subequations}
where:
\begin{itemize}
\item  $b \in \left\{ 1, \ldots, C \right\}, \;j\in \nu \left(b\right),\; i \in \left\{ 1, \ldots, K_j \right\},\;  f \in \left\{ f_1, \ldots f_F \right\}, \;  m \in \left\{ m_1, \ldots m_L \right\}, \; p \in \left\{ 1, \ldots \tau \right\}$
\item $x_{ib}^{jfm}$ is a binary frequency allocation variable that equals $1$ when the $i$-th user in the $j$-th group of cell $b$ is assigned frequency $f$ with modulation $m$, and $0$ otherwise.
\item $w_{ib}^j$ is a fairness weight associated with the $i$-th user belonging to the $j$th group in cell $b$. 
\item $R_m \in \left\{ R_{m_1}, \ldots, R_{m_L} \right\}$ represents the corresponding  bit rate of a given modulation scheme.
\item $\theta_m \in \left\{ \theta_{m_1}, \ldots, \theta_{m_L} \right\}$ is  the predefined SINR threshold associated with a given modulation.
\item $\mathcal{U}\left(j,b\right)$ denotes the set of indices of critical services users associated with the $j$-th group in cell $b$.
\item $\Pi\left(j,b,p\right)$ denotes the set of users allocated pilot $p$ within the $j$th group in cell $b$.
\item  The ${SINR}_{lr}^{kf}$ term (denoted here with indices $l,r,k$) is given by:
\begin{equation}
\begin{aligned}
    SINR_{lr}^{\left[ kf \right]} = &\frac{x_{lr}^{kfm}\left| Tr \left\{ \mathbf{C}_{lr}^{\left[ rk \right]} \mathbf{\Phi}_k \mathbf{\Phi}_k^H \right\} \right|^2}{\substack{x_{lr}^{kfm}Tr \left\{ \mathbf{C}_{lr}^{\left[ rk \right]} \mathbf{\Phi}_k \mathbf{\Phi}_k^H\mathbf{C}_{lr}^{\left[ rk \right]} \mathbf{\Phi}_k \mathbf{\Phi}_k^H\right\} \\ + \sum\limits_{\substack{b=1 \\ b \neq r}}^C\sum\limits_{ j \in \nu \left( b \right)} x_{lb}^{jfm}\mathcal{I}_{\text{inter},lr}^{[k|bj]}  + \sum\limits_{\substack{ j \in \nu \left( r\right) \\ j \neq k}}x_{lr}^{jfm}\mathcal{I}_{\text{intra},lr}^{[k|j]} \\ + \sum\limits_{b=1}^{C} \sum\limits_{j \in \nu \left( b \right)} \sum\limits_{\substack{i=1 \\ i \neq l}}^{\tau}x_{ib}^{jfm} \mathcal{I}_{\text{orth},lr}^{[k|bji]}\\+ \sum\limits_{b=1}^{C}\sum\limits_{j \in \nu \left( b \right)}\sum\limits_{i=1}^{\tau} x_{ib}^{jfm}\mathcal{I}_{\text{err},lr}^{[k|bji]} \\+ x_{lr}^{kfm} \sigma^2 Tr \left\{ \mathbf{C}_{lr}^{\left[ rk \right]} \mathbf{\Phi}_k \mathbf{\Phi}_k^H \right\}}} .
\end{aligned}
    \end{equation} 
\end{itemize}

\begin{rem}
To prioritize the strict requirements of critical services without unnecessarily allocating excess resources to them once their demands are met, the fairness weights are designed such that $w_{ib}^j = 0$ for critical services users and $w_{ib}^j = 1$ (or another positive priority weight) for standard users. Consequently, the objective function \eqref{mip:cout} effectively maximizes the sum-rate of standard best-effort traffic, while the required data rates for critical services are strictly guaranteed by the hard constraint \eqref{mip:urllc}.
\end{rem}

\begin{rem}
The deterministic constraints formulated above assume perfect knowledge of the true spatial channel covariance matrices ($\mathbf{R}$). Because the covariance matrix of the estimated channel ($\mathbf{C}$) is derived directly from $\mathbf{R}$, any inaccuracies in $\mathbf{R}$ propagate through the system. In practical deployments, the base station must estimate $\mathbf{R}$ via finite sample averaging, leading to imperfect covariance estimation. Such estimation errors induce a mismatch between the calculated and actual spatial leakage, increasing residual inter-group interference and degrading the true received SINR. To safeguard the strict reliability of critical services against this uncertainty, practical implementations typically introduce conservative SINR backoff margins. Extending this framework to robust optimization, which explicitly accounts for covariance estimation error bounds, is a compelling direction for future work.
\end{rem}

In optimization problem \eqref{Opt_proble_frerq_modul}, constraint \eqref{mip:1modulation} ensures that each scheduled user is assigned only one modulation per allocated frequency. Constraint \eqref{mip:tau} guarantees that at most $\tau$ users belonging to the same group in a given cell are assigned the same frequency $f$. Constraint \eqref{mip:sinr} enforces the SINR requirement for a scheduled user assigned modulation $m$ and frequency $f$, ensuring that the transmission satisfies a reliability (or error) threshold. Finally, constraint \eqref{mip:urllc} ensures a strict  transmit rate threshold  for scheduled critical services users. This constraint must be satisfied for critical services only, whereas the allocated resources for the other  non real time data applications is driven only by the maximization of the objective function in (24a). The use of weights $w^j_{i,b}$ allows fairness between these non-critical applications.  

\begin{rem}
While our formulation utilizes a deterministic SINR lower bound to ensure immediate feasibility for discrete modulation assignment, evaluating the exact outage probability for critical services requires analyzing the full statistical distribution of the SINR. Recent analytical studies, such as the derivation of the exact SINR probability density and cumulative distribution functions under pilot contamination in \cite{ChongSINR}, provide a rigorous mathematical foundation for such statistical analysis and represent a highly relevant direction for future extensions of this framework.\end{rem}

All the aforementioned constraints are linear except for \eqref{mip:sinr}, which is quadratic. We linearize it as follows. The constraint takes the general form:
\begin{eqnarray}
\frac{a_i y_i}{ \sum_{j \in I} b_j y_j } \geq y_i c_i ,   \forall i \in I,
\end{eqnarray}
where $y_i$ are boolean variables, $a_i, b_i$, and $c_i$ are strictly positive constants, and $I$ represents the set of indices. Provided that the validity condition $b_i \leq \frac{a_i}{c_i}$ holds, this is mathematically equivalent to the following linear constraint:
\begin{eqnarray}
 \sum_{j \in I} b_j y_j \leq \frac{a_i}{c_i} + (1-y_i) \sum_{j \in I\setminus \{i\}} b_j,  \forall i \in I.
\end{eqnarray}

In our specific framework, the condition $b_i \leq \frac{a_i}{c_i}$ ensures that a user's isolated SINR (considering only channel uncertainty and noise) is greater than or equal to the modulation threshold $\theta_m$. If this condition is violated, the user cannot support the modulation even without multi-user interference, rendering the assignment automatically infeasible. Indeed, if $y_i = 1$, the two inequalities are equivalent, and if $y_i = 0$, both constraints are automatically satisfied. Applying this transformation, constraint \eqref{mip:sinr} becomes:
\begin{equation}
    \substack{x_{lr}^{kfm}Tr \left\{ \mathbf{C}_{lr}^{\left[ rk \right]} \mathbf{\Phi}_k \mathbf{\Phi}_k^H\mathbf{C}_{lr}^{\left[ rk \right]} \mathbf{\Phi}_k \mathbf{\Phi}_k^H\right\}  + \sum\limits_{\substack{b=1 \\ b \neq r}}^C\sum\limits_{ j \in \nu \left( b \right)} x_{lb}^{jfm}\mathcal{I}_{\text{inter},lr}^{[k|bj]} \\ + \sum\limits_{\substack{ j \in \nu \left( r\right) \\ j \neq k}}x_{lr}^{jfm} \mathcal{I}_{\text{intra},lr}^{[k|j]}  + \sum\limits_{b=1}^{C} \sum\limits_{j \in \nu \left( b \right)} \sum\limits_{\substack{i=1 \\ i \neq l}}^{\tau}x_{ib}^{jfm}\mathcal{I}_{\text{orth},lr}^{[k|bji]} \\ + \sum\limits_{b=1}^{C}\sum\limits_{j \in \nu \left( b \right)}\sum\limits_{i=1}^{\tau} x_{ib}^{jfm}\mathcal{I}_{\text{err},lr}^{[k|bji]} + x_{lr}^{kfm} \sigma^2Tr \left\{ \mathbf{C}_{lr}^{\left[ rk \right]} \mathbf{\Phi}_k \mathbf{\Phi}_k^H \right\} \\
    \leq \frac{\left| Tr \left\{ \mathbf{C}_{lr}^{\left[ rk \right]} \mathbf{\Phi}_k \mathbf{\Phi}_k^H \right\} \right|^2}{\theta_m} +(1-x_{lr}^{kfm}) \bigg( \sum\limits_{\substack{b=1 \\ b \neq r}}^C\sum\limits_{ j \in \nu \left( b \right)}\mathcal{I}_{\text{inter},lr}^{[k|bj]}  \\ + \sum\limits_{\substack{ j \in \nu \left( r\right) \\ j \neq k}} \mathcal{I}_{\text{intra},lr}^{[k|j]}  + \sum\limits_{b=1}^{C} \sum\limits_{j \in \nu \left( b \right)} \sum\limits_{\substack{i=1 \\ i \neq l}}^{\tau} \mathcal{I}_{\text{orth},lr}^{[k|bji]} \\ + \sum\limits_{b=1}^{C}\sum\limits_{j \in \nu \left( b \right)}\sum\limits_{i=1}^{\tau} \mathcal{I}_{\text{err},lr}^{[k|bji]} -  \mathcal{I}_{\text{err},lr}^{[k|rkl]} \bigg)} .
\end{equation}
The optimization problem \eqref{Opt_proble_frerq_modul} is a mixed-integer linear programming (MILP) problem. In general, finding the global optimum for such problems using polynomial-time algorithms is NP-hard. Before proceeding to investigate the computational complexity of problem \eqref{Opt_proble_frerq_modul}, we first analyze a tractable special case.
\subsection{Two cells  Case} \label{sec3B}
We consider a two-cell scenario where only one group of $n=\tau$ users is scheduled per cell. Moreover, we assume there are $n$ frequencies and $L$ modulation schemes available for assignment. Furthermore, we assume that:
\begin{itemize}
\item 	$\theta_m > 1 , \, \forall m \in \left\{1,2, \ldots, L\right\}$. 
\item  $ \beta_{1b}^{\left[b\right]}>\beta_{2b}^{\left[b\right]}>\ldots>\beta_{nb}^{\left[b\right]}, \, \forall\ b \in \left\{1,2\right\}$.
\item $ \theta_m \beta_{nb}^{\left[b\right]} > \beta_{ib}^{\left[b\right]}, \, \forall i =\{ 1, \ldots, n-1\}, \, \forall\ b \in \left\{1,2\right\}$, $\forall\ m \in \left\{1,2, \ldots, L\right\}$.
\end{itemize}
Consequently, intra-cell interference is eliminated since each cell serves only one group of scheduled users, leaving only inter-cell interference. Therefore, problem \eqref{Opt_proble_frerq_modul} reduces to:
\begin{equation}\label{eq:specal_pro}
\begin{aligned}
\max_{\textbf{x}}~ & \sum\limits_{b=1}^{2}\sum\limits_{i=1}^{n}{w_{ib}\sum\limits_{f=1}^{n}\sum\limits_{m=1}^{L}{x_{ib}^{fm}R_m}} \\
\textrm{s.t.} ~& \sum\limits_{m=1}^{L}x_{ib}^{fm}  \leq 1, \; \forall b = \left\{ 1, 2 \right\},  \;  \forall i = \left\{ 1, \ldots, n \right\},\\ ~& \;\qquad\qquad\qquad\;  \forall f = \left\{ 1, \ldots n \right\},  \\
~& {SINR}_{ib}^{f}  \geq  x_{ib}^{fm}\theta_m, \; \forall b = \left\{ 1, 2 \right\}, \;  \forall i = \left\{ 1, \ldots, n \right\}, \\ ~&\qquad\qquad\qquad \forall f = \left\{ 1, \ldots n \right\},  \;\forall m = \left\{ 1, \ldots L \right\},
\end{aligned}
\end{equation}
where ${SINR}_{ib}^{f}$ is given by:

\begin{equation}
SINR_{ib}^{f}=\frac{\left( \beta_{ib}^{\left[ b\right]}\right)^2}{\sum\limits_{\substack{b^\prime=1 \\ b^\prime \neq b}}^2 \sum\limits_{j=1}^{n}{\left( \beta_{jb^\prime}^{\left[ b\right]}\right)^2x_{jb^\prime}^{fm}}} .
\end{equation}

\begin{theorem}\label{theorem:polynomialcase}
The optimization problem \eqref{eq:specal_pro} is polynomial-time solvable.
\end{theorem}
\begin{proof}
Please refer to Appendix \ref{Ap:proofPolyn} for more details. 
\end{proof}

\begin{rem}
It is important to note that the assumptions defining this special case are highly restrictive relative to the general problem \eqref{Opt_proble_frerq_modul}. These conditions pre-impose a structure where all SINR constraints reduce strictly to pairwise exclusivity. While this does not strictly require a near-degenerate modulation set, it does impose rigid inequalities between the large-scale fading coefficients $\beta$ and the modulation SINR thresholds $\theta_m$, serving primarily as a theoretical baseline to establish polynomial-time solvability under ideal bipartite conditions.
\end{rem}

\subsection{Study of NP-hardness}
In the previous subsection, we demonstrated that under specific conditions, the optimization problem \eqref{Opt_proble_frerq_modul} can be solved using a polynomial-time algorithm. However, when the number of cells is three or more ($C \geq 3$), finding the global optimum becomes computationally intractable. This is formalized in the following theorem.
\begin{theorem}\label{thm_nphard}
The optimization problem formulated in \eqref{Opt_proble_frerq_modul} is strongly NP-Hard for $C \geq 3$.
\label{theorem:nphard}
\end{theorem}
\begin{proof}
See Appendix \ref{AP:oproofNPhard}.
\hfill
\end{proof}

The complexity analysis results for problem \eqref{Opt_proble_frerq_modul} are summarized in Table \ref{tb:comple}.
\begin{table}[ht!]
\caption{Complexity Analysis of the Resource Allocation Problem}
\label{tb:comple}
\begin{tabularx}{\columnwidth}{|p{2.5cm}||X||X|} 
\hline
 & \textbf{Two cells} & \textbf{Three or more cells} \\
\hline
\textbf{One modulation} & \textit{(1)} Polynomial-Time Solvable & \textit{(2)} Strongly NP-Hard \\
\hline
\textbf{\textit{L} modulations} & \textit{(3)} Polynomial-Time Solvable\footnotemark & \textit{(4)} Strongly NP-Hard \\
\hline
\end{tabularx}
\end{table}
\footnotetext{Holds strictly under the ordered assumptions of the special case detailed in Section \ref{sec3B}.}

\begin{rem} 
Note that case (1) in Table \ref{tb:comple} is a special instance of case (3); therefore, the polynomial-time solvability of (3) inherently applies to (1). Similarly, because case (4) is a generalization of case (2), the optimization problem associated with multiple modulations is at least as difficult as the single-modulation case, meaning case (4) is also strongly NP-hard.
\end{rem}

\subsection{Suboptimal Solution Approach}
Given the strong NP-hardness of the general resource allocation problem \eqref{Opt_proble_frerq_modul}, we propose a computationally efficient, low-complexity algorithm based on linear programming (LP) relaxation and greedy rounding. The suboptimal solution is obtained by applying the following steps:
\begin{itemize}
    \item \textbf{LP Relaxation:} Relax the binary constraints on the allocation variables to continuous variables $x \in [0,1]$ and solve the resulting relaxed LP to obtain a fractional solution.
    \item \textbf{Critical Services Prioritization:} For all critical services users, greedily select the highest fractional values of $x$ and round them to $1$ until their specific rate demands are fully satisfied, thereby guaranteeing strict reliability.
    \item \textbf{Standard User Allocation:} For the remaining standard users, apply the same greedy rounding strategy to the highest fractional values of $x$, rounding them to $1$ while ensuring all constraints are respected.
\end{itemize}

\textit{Computational Complexity Analysis:} 
The theoretical complexity of the proposed suboptimal algorithm is evaluated in two parts: the LP relaxation and the greedy rounding step. Let $V$ denote the total number of decision variables, given by $V = \sum_{b=1}^{C} \sum_{j \in \nu(b)} K_j \times F \times L$, where $K_j \leq \tau$. Solving the relaxed linear program using standard interior-point methods incurs a worst-case polynomial complexity of $\mathcal{O}(V^3)$. Subsequently, the greedy rounding step requires sorting the fractional values, which has a complexity of $\mathcal{O}(V \log V)$, followed by a sequential iteration to finalize the resource assignments in $\mathcal{O}(V)$ time. Therefore, the overall computational complexity of the proposed approach is strictly bounded by $\mathcal{O}(V^3)$. This polynomial-time complexity represents a massive reduction compared to the exact MILP solver (used for the global optimum), which entails a worst-case exponential complexity of $\mathcal{O}(2^V)$.

The performance of this proposed approach is evaluated numerically in Section \ref{sec:numericalresults}.

\section{Numerical Results}\label{sec:numericalresults}
In this section, we evaluate the proposed approach via Monte Carlo simulations averaged over 5000 independent realizations (50 cell geometries $\times$ 100 channel draws). We consider single-cell ($C=1$) and multi-cell ($C=3$, inter-site distance $300$\,m) systems. Each BS is equipped with a uniform linear array of $M=32$ antennas (half-wavelength spacing) and serves $K=8$ single-antenna users. Users are uniformly distributed in a $120^\circ$ sector at distances $r_k \in [100,150]$\,m. Large-scale fading follows a power-law pathloss model with exponent $3$ and reference coefficient $10^{-2.6}$. Small-scale fading is modeled via a correlated Rayleigh channel (one-ring model \cite{shiu}) projected into the beam domain. Beam-domain clustering uses an $M\times M$ DFT matrix with cluster size $d=8$, yielding $G=4$ groups per BS.
System parameters are set as follows: pilot length $\tau = 3$, pilot power $\rho_{p}=1$, noise variance $\sigma^2 = 3.4\times 10^{-12}$, and $F = 5$ frequency resources. The scheduler utilizes three modulation formats with spectral efficiencies $R_m \in \{2,4,6\}\,\text{bps/Hz}$ and linear SINR thresholds $\theta_{\mathrm{mod}} \in \{0.9,1.8,4\}$. 
Based on the path-loss parameters and user distances, the resulting nominal single-antenna operating SNR ranges from approximately $23.4$ dB at the cell edge to $28.7$ dB at the cell center. To model dynamic traffic, each user initiates critical services traffic with probability $p_{\mathrm{URLLC}} = 0.2$, drawing a rate requirement $R_{\mathrm{URLLC}}$ uniformly from $[10,15]\,\text{bps/Hz}$. Finally, to ensure fairness and prevent the objective function from unnecessarily maximizing critical services rates beyond their strict requirements, the fairness weights are assigned as $w=0$ for critical services users and $w=1$ for standard users.

It is important to emphasize the rationale behind the chosen performance metrics. In the proposed framework, the strict latency and reliability requirements of critical services are translated into deterministic minimum rate and SINR constraints (as defined in \eqref{mip:sinr} and \eqref{mip:urllc}). Therefore, for any feasible resource allocation solution generated by the algorithms, these critical service requirements are strictly met by design. Consequently, to comprehensively evaluate system performance, our numerical analysis focuses on two complementary dimensions. First, for feasible network states, we evaluate the system sum-rate to demonstrate coexistence efficiency—how well the proposed approaches maximize the throughput of standard best-effort traffic while strictly safeguarding critical services. Second, we evaluate the probability of feasibility to provide insight into overall network dimensioning and capacity limits as critical traffic loads increase and resources approach exhaustion.

\subsection{Single Cell analysis}
\begin{figure}[!t]
\centering
\includegraphics[width=2.9in]{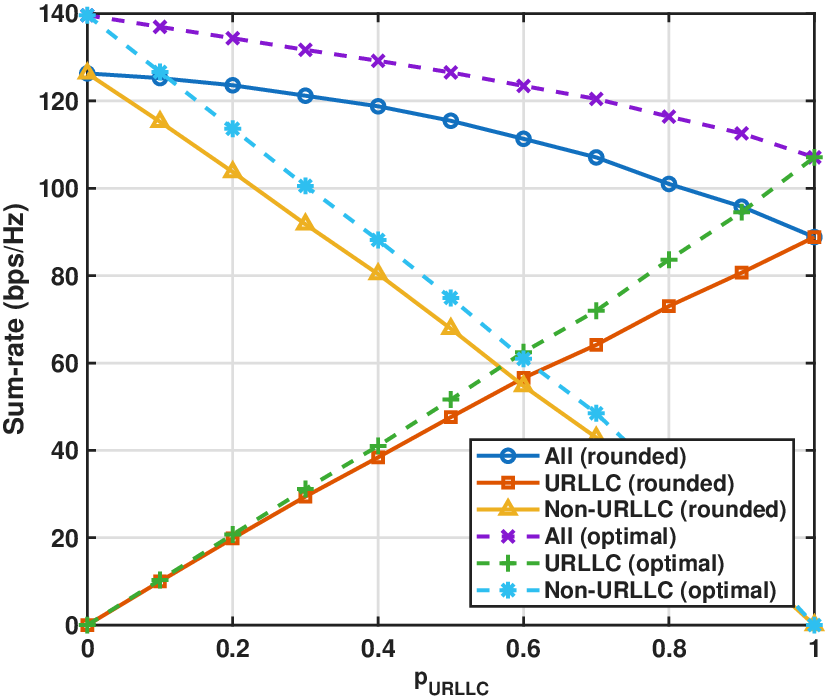}
\caption{Average sum bit rate in bits/s/Hz in terms of $p$ with $C=1$ and $\alpha=0.55$.}
\label{fig:bitratpURLLC}
\end{figure}

Fig. \ref{fig:bitratpURLLC} illustrates the impact of the critical services traffic probability, $p_{\mathrm{URLLC}}$, on the system performance. As $p_{\mathrm{URLLC}} $ increases, there is a clear decrease in the total system sum-rate. This reduction occurs because the scheduler must prioritize the strict rate and reliability constraints of critical services users, often necessitating the allocation of resources that would otherwise be used for high-throughput best-effort traffic. Consequently, while the critical services sum-rate grows almost linearly with $p_{\mathrm{URLLC}} $, the average sum bit rate of non-critical services traffic drops sharply as these users are increasingly displaced to accommodate the urgent traffic. Furthermore, the results establish the efficiency of the proposed rounding algorithm; the solid lines representing the rounded solutions closely track the dashed lines of the optimal MILP results across the entire range of $p_{\mathrm{URLLC}}$, demonstrating that a near-optimal resource allocation is achieved with a minimal optimality gap.

\begin{figure}[!t]
\centering
\includegraphics[width=2.9in]{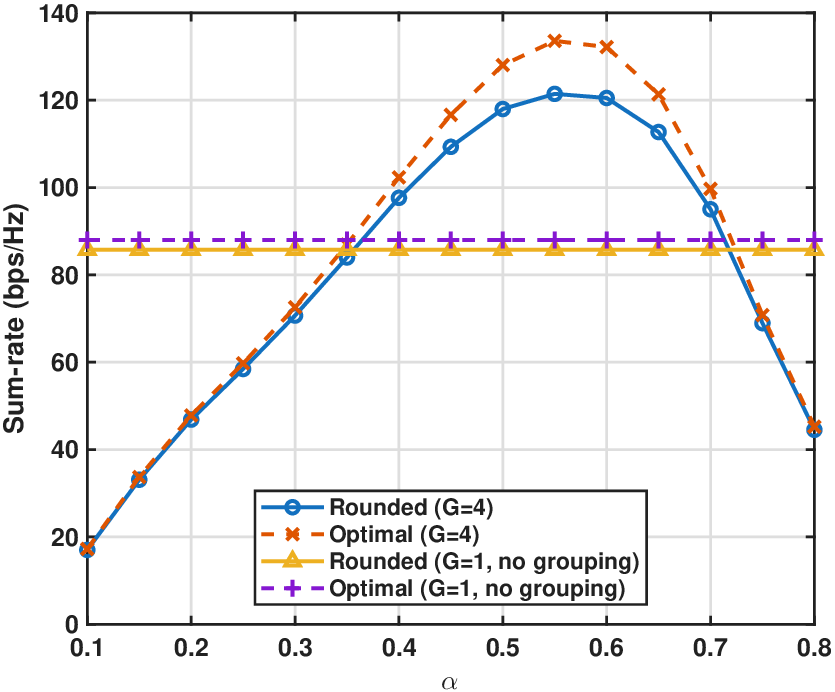}
\caption{Average sum bit rate in bits/s/Hz in terms of $\alpha$ with $C=1$.}
\label{fig:bitratealpha}
\end{figure}

Fig. \ref{fig:bitratealpha} portrays the performance evaluation in terms of sum throughput versus the threshold $\alpha$. The results indicate that the spatial grouping approach ($G=4$) significantly outperforms the standard mMIMO baseline without grouping ($G=1$), provided $\alpha$ is appropriately tuned. We observe that the sum-rate exhibits a concave behavior with respect to $\alpha$. For low values of $\alpha$, the system is highly selective, which minimizes intra-cell interference but also limits the number of users who can satisfy the strict clustering condition. As $\alpha$ increases, more users are successfully assigned to clusters, leading to a substantial increase in the total throughput. However, beyond a certain threshold (approximately $\alpha > 0.7$), the overall throughput collapses. This collapse occurs because at high values of $\alpha$, the user's channel energy is dispersed between several groups. Consequently, users fail to isolate a unique dominant group that satisfies the clustering criterion in \eqref{eq:inegalpha}. Because no single group can be uniquely assigned, the grouping stage fails for these users. They are subsequently dropped from the scheduling pool and the resource allocation optimizer never even considers them, leading to a sharp drop in system sum rate. This highlights the critical importance of carefully tuning the threshold $\alpha$ to balance spatial isolation with successful group attribution, thereby avoiding severe user dropout. Finally, the rounding algorithm maintains a marginal optimality gap compared to the MILP solution, confirming its practical effectiveness. As visually evident, the suboptimal solution strictly lower-bounds the optimal baseline, coinciding exactly in regions where the relaxed linear program naturally yields integer or near-integer solutions.

\begin{figure}[!t]
\centering
\includegraphics[width=2.9in]{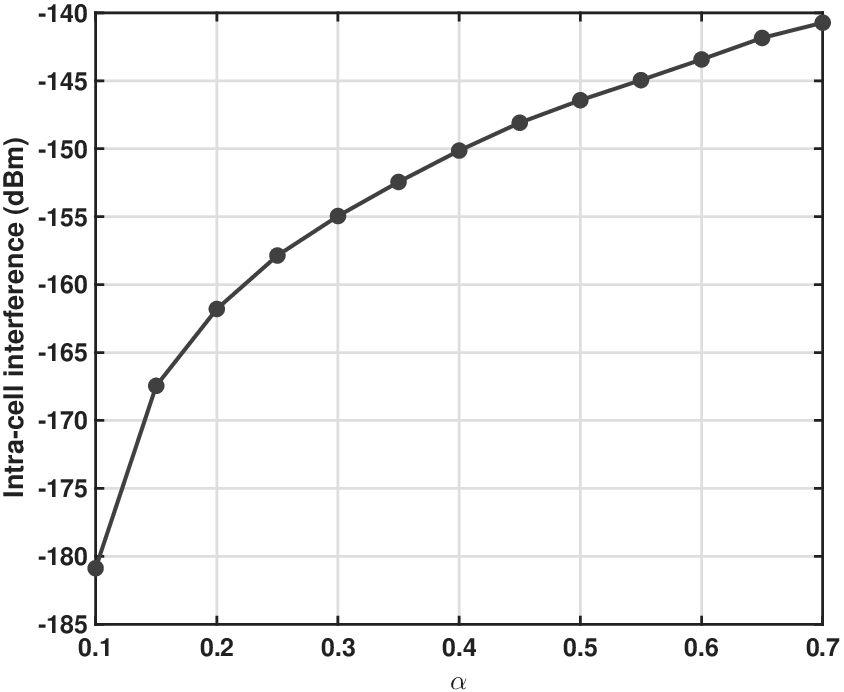}
\caption{Average intra-cell interference in terms of $\alpha$ with $C=1$.}
\label{fig:intra}
\end{figure}

Fig. \ref{fig:intra} illustrates the evolution of the average intra-cell interference (in dBm) as a function of the clustering threshold $\alpha$. The numerical results corroborate the theoretical analysis presented in Lemma \ref{lemme:sinr}, where the intra-cell interference terms are scaled by a factor of $\alpha^4$. Because the physical large-scale channel gains $\beta$ are on the order of $10^{-9}$, the resulting interference powers ($\beta^2$) naturally reside near $-150$ dBm for strict clustering thresholds. As $\alpha$ increases, the spatial selectivity of the clusters relaxes, causing the intra-cell interference to grow strictly in accordance with the $\alpha^4$ polynomial scaling. Notably, this growth remains relatively slow for $\alpha < 0.4$, after which it exhibits a sharp non-linear increase as the clustering condition in \eqref{eq:inegalpha} becomes more permissive. This behavior demonstrates that the proposed grouping strategy successfully confines user channels to near-orthogonal subspaces, effectively making intra-cell interference a negligible component of the total system interference, provided $\alpha$ is maintained within a reasonable operating range.

\subsection{Multi-Cell analysis}
\begin{figure}[!t]
\centering
\includegraphics[width=2.9in]{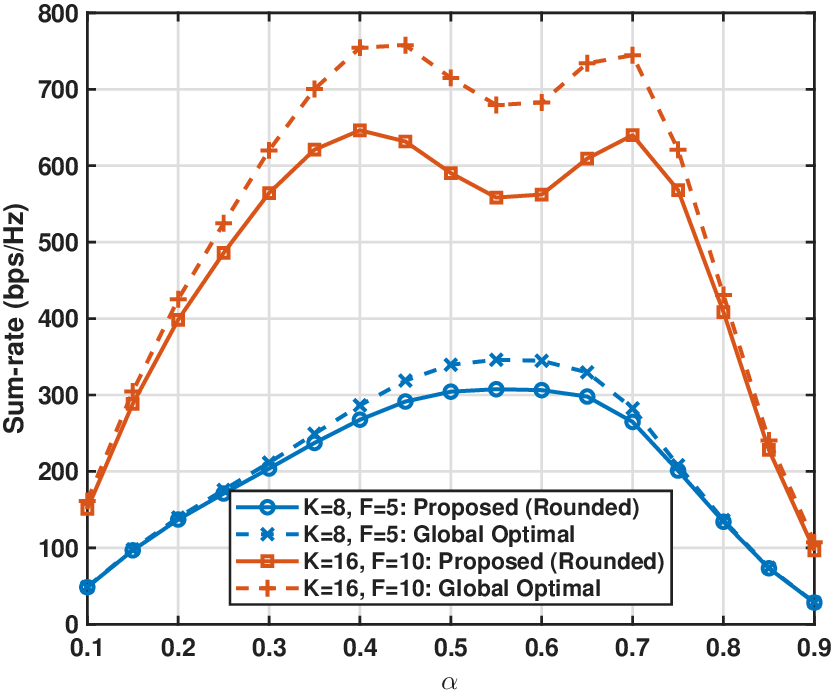}
\caption{Average sum bit rate in bits/s/Hz in terms of $\alpha$ with $C=3$.}
\label{fig:bitratealphaC3}
\end{figure}

In this subsection, we suppose that the network has 3 base stations each equipped with $32$ antennas. To demonstrate scalability, we analyze both a baseline configuration ($K=8$ users, $F=5$ frequencies) and a dense deployment ($K=16$ users, $F=10$ frequencies). The solutions for the frequency and modulation allocation are computed as in Section \ref{sec:frequencyallocation}. Fig. \ref{fig:bitratealphaC3} illustrates the total network throughput in a multi-cell environment ($C=3$) as a function of the clustering threshold $\alpha$. For the $K=8$ case, similar to the single-cell scenario, the system performance exhibits a concave relationship with the threshold, reaching a peak sum-rate of approximately 340 bps/Hz at $\alpha \approx 0.55$. At low values of $\alpha$, the throughput is significantly degraded due to overly strict cluster allocation. In contrast, the scaled $K=16$ network displays a distinct non-concave behavior with two local maxima. It first peaks at $\alpha = 0.45$ (achieving $\sim 757$ bps/Hz), where strict spatial isolation perfectly balances the denser user load. As the threshold becomes more permissive, user channel power may satisfy the condition in \eqref{eq:inegalpha} for multiple reference beams, leading to assignments in clusters with poor spatial alignment and high inter-cluster leakage. This causes a temporary dip in the $K=16$ throughput, before a secondary peak emerges near $\alpha = 0.7$ ($\sim 744$ bps/Hz) as the algorithm aggressively multiplexes a larger subset of the dense user pool across the expanded frequency resources. Conversely, as $\alpha$ increases beyond these optimal points, the overall throughput collapses sharply. This collapse is not caused by the spatial separation requirement becoming stricter, but rather by the spatial energy becoming too dispersed among candidate groups. Without a unique group assignment, the grouping process fails. These users are dropped before the resource allocation optimizer is even executed, causing the total throughput to plummet. Once again, this underscores the necessity of selecting an optimal $\alpha$ to prevent massive user dropout. Furthermore, the results highlight that the proposed rounding algorithm maintains a small optimality gap compared to the MILP solution, effectively capturing both the concave and non-concave trends while providing a computationally efficient alternative for dense multi-cell resource allocation.

\begin{figure}[!t]
\centering
\includegraphics[width=2.9in]{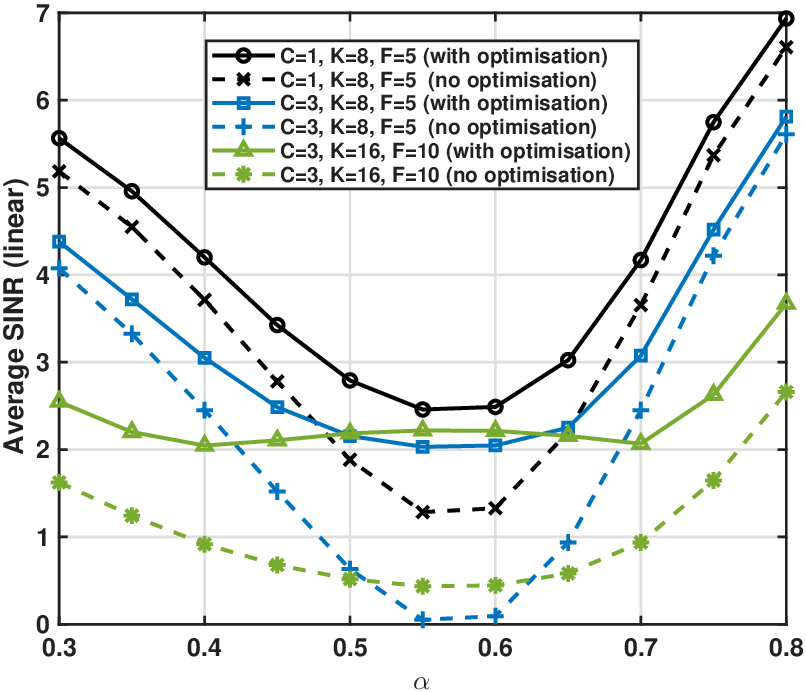}
\caption{Average SINR of all clustered users in terms of $\alpha$.}
\label{fig:sinr}
\end{figure}

Fig.~\ref{fig:sinr} illustrates the average achieved SINR as a function of the clustering threshold $\alpha$, comparing the performance of the spatial grouping strategy ($G=4$) with and without the proposed joint frequency and modulation allocation optimization for the single-cell scenario ($C=1, K=8, F=5$), as well as both the baseline ($C=3, K=8, F=5$) and dense ($C=3, K=16, F=10$) multi-cell scenarios. The results highlight that the resource allocation process significantly enhances the average SINR across the entire range of $\alpha$ for all configurations. Notably, the SINR gain provided by the optimization is most pronounced in the dense $K=16$ deployment. In this crowded scenario, unstructured transmission (no optimization) yields severe inter-user collisions, causing the average linear SINR to plummet to approximately $0.44$ at $\alpha \approx 0.55$. This demonstrates that while multi-cell environments are subject to severe inter-cell interference, the proposed optimization successfully manages these heightened interference levels by intelligently isolating users across orthogonal resource blocks and discrete modulations. As a result, the optimizer maintains the average linear SINR strictly above $2$ even in the most interference-heavy regions to satisfy the SINR requirements $\theta_m$. Interestingly, as $\alpha$ initially increases from low values, a decrease in SINR is observed due to the rising intra-cell and inter-cell interference caused by more permissive clustering. However, beyond $\alpha \approx 0.6$ (and $\alpha \approx 0.7$ for $K=16$), the average SINR rises sharply for all configurations. This paradoxical rise in average SINR at high $\alpha$ is a direct result of the grouping stage failures. As $\alpha$ becomes too high, the vast majority of users experience dispersed energy across multiple groups, fail to obtain a unique group assignment, and are dropped before the optimization begins. The optimizer only receives and processes the very small subset of users whose channels are exceptionally aligned with a single group. Therefore, the average SINR of these few scheduled users artificially inflates, even as the total system throughput simultaneously declines due to massive user dropout.

\begin{figure}[!t]
    \centering
    \includegraphics[width=2.9in]{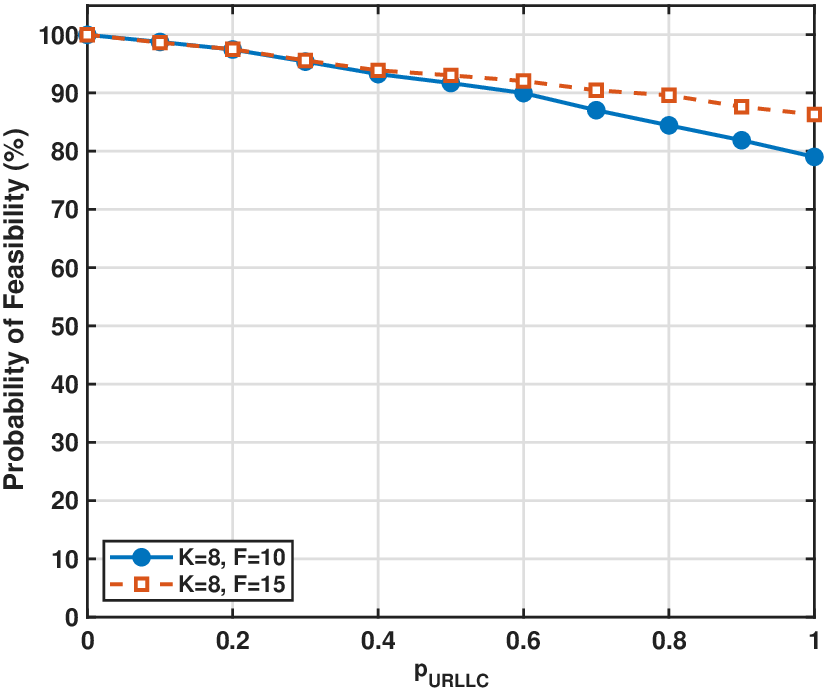}
\caption{Probability of feasibility, evaluated across 5000 independent realizations, vs. the critical traffic probability ($p_{\text{URLLC}}$) for a multi-cell network ($C=3, K=8$), comparing resource allocations of $F=10$ and $F=15$ blocks with a fixed pilot overhead ($\tau=3$).}
    \label{fig:feasibilityrate}
\end{figure}

Finally, Fig. \ref{fig:feasibilityrate} provides an insight into network dimensioning by plotting the feasibility of the optimization problem as a function of the critical traffic probability ($p_{\text{URLLC}}$). Since minimum rate and SINR constraints must be strictly met by the critical services, having a high proportion of critical services may exceed the overall network capacity and require more resources. As Fig. \ref{fig:feasibilityrate} clearly shows, network dimensioning and call admission control procedures should be applied in order to filter and admit users for which the problem is feasible. Although this goes beyond the scope of this paper, which deals mainly with analyzing and solving the optimization framework for the admitted users, developing these corresponding call admission control mechanisms represents a good direction for future research. In this evaluation, we focus on the multi-cell scenario ($C=3$) with a fixed user load of $K=8$ users per cell, comparing two physical resource configurations: $F=10$ and $F=15$ available resource blocks. Crucially, the number of orthogonal pilot sequences is kept strictly constant at $\tau=3$ in both cases to respect the limited channel coherence interval. As $p_{\text{URLLC}}$ increases, the feasibility declines because the aggregate critical data demands exhaust the finite available resource blocks. Expanding the available resources to $F=15$ mitigates this capacity bottleneck, allowing the system to robustly handle much higher critical traffic intensities before call admission control intervention becomes necessary.

To rigorously quantify the performance of the proposed suboptimal solution in multi-cell deployments, we evaluated the empirical optimality gap and computational complexity over the $5000$ independent realizations at $\alpha = 0.6$ and $p_{\text{URLLC}} = 0.2$. For the baseline scenario ($C=3, K=8, F=5$), the LP-rounding algorithm achieves a mean relative optimality gap of $8.77\%$ and a worst-case gap of $34.21\%$, requiring an average runtime of $0.0647$ seconds compared to $2.2156$ seconds for the exact MILP solver ($ \sim 34\times$ speedup). In the scaled, denser deployment ($C=3, K=16, F=10$), the exact MILP solver’s runtime surges to an average of $118.59$ seconds due to exponential combinatorial growth, whereas the proposed heuristic maintains a low runtime of only $0.48$ seconds, achieving a massive speedup factor of over $249\times$ while keeping the mean and worst-case optimality gaps tightly bounded at $13.31\%$ and $42.70\%$, respectively. This confirms that the rounding algorithm provides a highly scalable, low-latency resource allocation mechanism ideally suited for dense, latency-critical cellular networks.

\section{Conclusion}
\label{sec:conclusion}
In this paper, we investigated the coexistence of critical services and standard traffic in TDD mMIMO systems. To mitigate pilot contamination,  a spatial multiplexing framework that clusters users based on channel directions is used to effectively marginalizing intra-cell interference. To address residual inter-cell interference and guarantee strict critical services requirements, we formulated a joint frequency and modulation allocation problem. We theoretically proved that this optimization problem is strongly NP-hard. Consequently, we developed a low-complexity LP-relaxation rounding algorithm to efficiently approximate the solution. Numerical results demonstrate that our approach achieves near-optimal sum-rates while strictly safeguarding critical services reliability.

 \appendices 
 
 \setcounter{equation}{0}
\renewcommand{\theequation}{A.\arabic{equation}}      
\section{Proof of Lemma \ref{lemme:sinr}}\label{ProofSinr}
For any arbitrary cluster size $d$, the detected symbol for the $l$-th user in the $k$-th group of cell $r$ can be expanded as:
\begin{align}
    {\hat{s}}_{klr} & = \sum\limits_{b=1}^{C}\sum\limits_{j\in\nu\left(b \right )}\sum\limits_{i=1}^{\tau}{\hat{\mathbf{g}}_{lr}^{[rk]H}}\boldsymbol{\Phi }_k\boldsymbol{\Phi }_k^H\boldsymbol{\mathbf{g}}_{ib}^{[rj]}s_{jib} \nonumber\\ &\qquad+{\hat{\mathbf{g}}_{lr}^{[rk]H}}\boldsymbol{\Phi }_k\boldsymbol{\Phi }_k^H\boldsymbol{\mathbf{n}}_u  \\
    & =  \underbrace{{\hat{\mathbf{g}}_{lr}^{[rk]H}}\boldsymbol{\Phi }_k\boldsymbol{\Phi }_k^H\boldsymbol{\mathbf{n}}_u}_{I_5} \nonumber\\& \quad + \underbrace{\sum\limits_{b=1}^{C}\sum\limits_{j\in\nu\left(b \right )}\sum\limits_{i=1}^{\tau}{\hat{\mathbf{g}}_{lr}^{[rk]H}}\boldsymbol{\Phi }_k\boldsymbol{\Phi }_k^H\tilde{\boldsymbol{\mathbf{g}}}_{ib}^{[rj]}s_{jib}}_{I_4} \nonumber \\ & \quad + \sum\limits_{b=1}^{C}\sum\limits_{j\in\nu\left(b \right )}\sum\limits_{i=1}^{\tau}{\hat{\mathbf{g}}_{lr}^{[rk]H}}\boldsymbol{\Phi }_k\boldsymbol{\Phi }_k^H\hat{\boldsymbol{\mathbf{g}}}_{ib}^{[rj]}s_{jib} \\ 
    & =  I_5 + I_4 \nonumber \\& \quad + \sum\limits_{b=1}^{C}\sum\limits_{j\in\nu\left(b \right )}\Bigg( \sum\limits_{\substack{{i=1} \\ i\neq l  }}^{\tau}{\hat{\mathbf{g}}_{lr}^{[rk]H}\boldsymbol{\Phi }_k\boldsymbol{\Phi }_k^H\hat{\boldsymbol{\mathbf{g}}}_{ib}^{[rj]}s_{jib}}\nonumber  \\& \quad+ {\hat{\mathbf{g}}_{lr}^{[rk]H}}\boldsymbol{\Phi }_k\boldsymbol{\Phi }_k^H\hat{\boldsymbol{\mathbf{g}}}_{lb}^{[rj]}s_{jlb} \Bigg) \\
    & =  I_5 + I_4 \nonumber \\& \quad 
    + \underbrace{\sum\limits_{b=1}^{C}\sum\limits_{j\in\nu\left(b \right )}\sum\limits_{\substack{{i=1} \\ i\neq l  }}^{\tau}{\hat{\mathbf{g}}_{lr}^{[rk]H}}\boldsymbol{\Phi }_k\boldsymbol{\Phi }_k^H\hat{\boldsymbol{\mathbf{g}}}_{ib}^{[rj]}s_{jib}}_{I_3} \nonumber\\ & \quad  + \sum\limits_{b=1}^{C}\sum\limits_{j\in\nu\left(b \right )}{\hat{\mathbf{g}}_{lr}^{[rk]H}}\boldsymbol{\Phi }_k\boldsymbol{\Phi }_k^H\hat{\boldsymbol{\mathbf{g}}}_{lb}^{[rj]}s_{jlb} \\
    & =  I_5 + I_4 + I_3 + \underbrace{\sum\limits_{b=1}^{C}\sum\limits_{\substack{j\in\nu\left(b \right ) \\ jb \neq kr}}{\hat{\mathbf{g}}_{lr}^{[rk]H}}\boldsymbol{\Phi }_k\boldsymbol{\Phi }_k^H\hat{\boldsymbol{\mathbf{g}}}_{lb}^{[rj]}s_{jlb}}_{I_2} \nonumber\\
    & \; + \underbrace{{\hat{\mathbf{g}}_{lr}^{[rk]H}}\boldsymbol{\Phi }_k\boldsymbol{\Phi }_k^H\hat{\boldsymbol{\mathbf{g}}}_{lr}^{[rk]}s_{klr} - s_{klr} \mathop{\mathbb{E}}\left [ {\hat{\mathbf{g}}_{lr}^{[rk]H}}\boldsymbol{\Phi }_k\boldsymbol{\Phi }_k^H\hat{\boldsymbol{\mathbf{g}}}_{lr}^{[rk]} \right ] }_{I_1} \nonumber\\ & \quad +  \underbrace{s_{klr} \mathop{\mathbb{E}}\left [ {\hat{\mathbf{g}}_{lr}^{[rk]H}}\boldsymbol{\Phi }_k\boldsymbol{\Phi }_k^H\hat{\boldsymbol{\mathbf{g}}}_{lr}^{[rk]} \right ] }_{I_0}  \\
    & =  I_0 + I_1 + I_2 + I_3 + I_4 + I_5 .
\end{align}
Here, $\mathbb{E}[\cdot]$ denotes the conditional expectation given that the user satisfies the clustering condition. The expanded terms $I_0, \ldots, I_5$ are mutually uncorrelated. After applying the matched filter, the effective channel gain is given by $\mathop{\mathbb{V}\text{ar}}(I_0)$, which depends solely on the statistical properties of the channel. The effective interference and noise variance is simply the sum of the variances of the remaining terms. To derive the finite-regime uplink SINR, we resort to an approach similar to \cite{salah,ngo}, although the SINR expression is different in this work and requires more derivations.  We independently compute $\mathop{\mathbb{V}\text{ar}}(I_i)$ for $i=0,\ldots,5$ as follows:

The term $I_0$ represents the \textbf{effective channel gain} for the detected symbol $s_{klr}$, given by:
\begin{align}
\mathop{\mathbb{V}ar}\left(I_0\right)&=\mathbb{V}ar\left(s_{klr}\ \mathbb{E}\left[{{\hat{\mathbf{g}}}_{lr}^{\left[rk\right]H}}\mathbf{\Phi}_k{\mathbf{\Phi}_k}^H{\hat{\mathbf{g}}}_{lr}^{\left[rk\right]}\right]\right) \\
&=\left|\mathbb{E}\left[{{\hat{\mathbf{g}}}_{lr}^{\left[rk\right]H}}\mathbf{\Phi}_k{\mathbf{\Phi}_k}^H{\hat{\mathbf{g}}}_{lr}^{\left[rk\right]}\right]\right|^2 .
\end{align}
Applying the trace identity, we compute:
\begin{align}
\mathop{\mathbb{E}}\left[{{\hat{\mathbf{g}}}_{lr}^{\left[rk\right]H}}\mathbf{\Phi}_k{\mathbf{\Phi}_k}^H{\hat{\mathbf{g}}}_{lr}^{\left[rk\right]}\right]&=\mathbb{E}\left[Tr\left({\hat{\mathbf{g}}}_{lr}^{\left[rk\right]}{{\hat{\mathbf{g}}}_{lr}^{\left[rk\right]H}}\mathbf{\Phi}_k{\mathbf{\Phi}_k}^H\right)\right] \\
&=Tr\left\{\mathbb{E}\left[{\hat{\mathbf{g}}}_{lr}^{\left[rk\right]}{{\hat{\mathbf{g}}}_{lr}^{\left[rk\right]H}}\right]\mathbf{\Phi}_k{\mathbf{\Phi}_k}^H\right\} \\
&= Tr\left\{\mathbf{C}_{lr}^{[rk]}\mathbf{\Phi}_k{\mathbf{\Phi}_k}^H\right\} .
\end{align}
Consequently, the variance becomes:
\begin{equation}
    \mathop{\mathbb{V}ar}\left(I_0\right) =\left | Tr\left\{\mathbf{C}_{lr}^{[rk]}\mathbf{\Phi}_k{\mathbf{\Phi}_k}^H\right\}   \right |^{2}.
\end{equation} 
The term $I_1$ accounts for \textbf{channel uncertainty}. Because the user relies on the statistical expectation of the effective channel rather than its instantaneous realization, this deviation acts as interference. Given that the data symbols $s_{klr}$ are mutually independent across users, we obtain:
\begin{align}
     &\mathop{\mathbb{V}ar}\left(I_1\right) \nonumber\\  = &\mathbb{V}ar  \Bigg( \hat{\mathbf{g}}_{lr}^{\left[rk\right]H}\mathbf{\Phi}_k{\mathbf{\Phi}_k}^H{\hat{\mathbf{g}}}_{lr}^{\left[rk\right]}s_{klr} \nonumber\\ &\qquad\qquad- s_{klr}\mathop{\mathbb{E}}\left[{{\hat{\mathbf{g}}}_{lr}^{\left[rk\right]H}}\mathbf{\Phi}_k{\mathbf{\Phi}_k}^H{\hat{\mathbf{g}}}_{lr}^{\left[rk\right]}\right] \Bigg) \\
    & = \mathbb{E}\left[s_{klr}{s_{klr}}^H\right] \nonumber\\ & \quad \times
    \mathbb{E}\left[ \left( \hat{\mathbf{g}}_{lr}^{\left[rk\right]H} {\mathbf{\Phi}}_k{\mathbf{\Phi}}_k^H\hat{\mathbf{g}}_{lr}^{\left[rk\right]}- \mathbb{E}\biggl[\hat{\mathbf{g}}_{lr}^{\left[rk\right]H} {\mathbf{\Phi}}_k{\mathbf{\Phi}}_k^H\hat{\mathbf{g}}_{lr}^{\left[rk\right]}\right]\right) \nonumber\\
    & \quad\> \times \left( \hat{\mathbf{g}}_{lr}^{\left[rk\right]H} {\mathbf{\Phi}}_k{\mathbf{\Phi}}_k^H\hat{\mathbf{g}}_{lr}^{\left[rk\right]}- \mathbb{E}\left[\hat{\mathbf{g}}_{lr}^{\left[rk\right]H} {\mathbf{\Phi}}_k{\mathbf{\Phi}}_k^H\hat{\mathbf{g}}_{lr}^{\left[rk\right]}\right]\right)^H \biggr] \\
    & =  \mathbb{E}\left[ \left| \hat{\mathbf{g}}_{lr}^{\left[rk\right]H} {\mathbf{\Phi}}_k{\mathbf{\Phi}}_k^H\hat{\mathbf{g}}_{lr}^{\left[rk\right]} \right| ^2 \right] - \left| \mathbb{E}\left[ \hat{\mathbf{g}}_{lr}^{\left[rk\right]H} {\mathbf{\Phi}}_k{\mathbf{\Phi}}_k^H\hat{\mathbf{g}}_{lr}^{\left[rk\right]} \right] \right| ^2 \\
    & = \mathop{\mathbb{V}ar}\left( \hat{\mathbf{g}}_{lr}^{\left[rk\right]H} {\mathbf{\Phi}}_k{\mathbf{\Phi}}_k^H\hat{\mathbf{g}}_{lr}^{\left[rk\right]} \right) \\
    & = Tr \left\{ \mathbf{C}_{lr}^{\left[ rk \right]} {\mathbf{\Phi}}_k{\mathbf{\Phi}}_k^H \mathbf{C}_{lr}^{\left[ rk \right]} {\mathbf{\Phi}}_k{\mathbf{\Phi}}_k^H  \right\} .
\end{align}
The last equality follows from Lemma \ref{lemma:trace_identity}.

The term $I_2$ captures the \textbf{pilot contamination effect}. Because pilot sequences are reused across cells and groups, the channel estimates $\hat{\mathbf{g}}_{lr}^{\left[rk\right]H}$ and ${{\hat{\mathbf{g}}}_{lb}^{\left[rj\right]H}}$ become correlated. This fundamental bottleneck of mMIMO persists even in the asymptotic antenna regime. Although impossible to eliminate entirely, our proposed spatial clustering approach significantly mitigates this interference.
\begin{align}
    &\mathop{\mathbb{V}ar}\left(I_2\right)  =  \mathop{\mathbb{V}ar} \left( \sum\limits_{b=1}^{C}\sum\limits_{\substack{j \in \nu \left( b \right) \\ jb \neq kr}}{  {\hat{\mathbf{g}}}_{lr}^{\left[ rk \right] H}\mathbf{\Phi}}_k \mathbf{\Phi}_k^H {\hat{\mathbf{g}}}_{lb}^{\left[ rj \right] }  s_{jlb}\right) \\
    & =  \sum\limits_{b=1}^{C} \sum\limits_{\substack{j \in \nu \left( b \right) \\  jb \neq kr}} \mathop{\mathbb{E}} \left[ s_{jlb} s_{jlb}^H \right] \nonumber\\&\quad \times\mathop{\mathbb{E}} \left[ \hat{\mathbf{g}}_{lr}^{\left[ rk \right] H} \mathbf{\Phi}_k \mathbf{\Phi}_k^H \hat{\mathbf{g}}_{lb}^{\left[ rj \right]}\hat{\mathbf{g}}_{lb}^{\left[ rj \right]H}\mathbf{\Phi}_k \mathbf{\Phi}_k^H \hat{\mathbf{g}}_{lr}^{\left[ rk \right] } \right] \\
    & =  \sum\limits_{\substack{b=1 \\ b \neq r}}^{C}\sum\limits_{j \in \nu \left( b\right)} \underbrace{\mathop{\mathbb{E}} \left[ \hat{\mathbf{g}}_{lr}^{\left[ rk \right] H} \mathbf{\Phi}_k \mathbf{\Phi}_k^H \hat{\mathbf{g}}_{lb}^{\left[ rj \right]}\hat{\mathbf{g}}_{lb}^{\left[ rj \right]H}\mathbf{\Phi}_k \mathbf{\Phi}_k^H \hat{\mathbf{g}}_{lr}^{\left[ rk \right] } \right]}_A \nonumber\\
    & \quad + \sum\limits_{\substack{j \in \nu \left( r\right) \\ j \neq k}}\mathop{\mathbb{E}} \left[ \hat{\mathbf{g}}_{lr}^{\left[ rk \right] H} \mathbf{\Phi}_k \mathbf{\Phi}_k^H \hat{\mathbf{g}}_{lr}^{\left[ rj \right]}\hat{\mathbf{g}}_{lr}^{\left[ rj \right]H}\mathbf{\Phi}_k \mathbf{\Phi}_k^H \hat{\mathbf{g}}_{lr}^{\left[ rk \right] } \right] .
\end{align}
Knowing that
\begin{equation}
    \hat{\mathbf{g}}_{lr}^{\left[rk\right]} = \mathbf{R}
    _{lr}^{\left[rk \right]} \left( \mathbf{I}_{M} + \sum\limits_{u=1}^{C}\sum\limits_{v \in \nu \left( u \right)} \mathbf{R}_{lu}^{\left[ rv \right]} \right)^{-1}\mathbf{Y}_p^{\left[ r \right]} \mathbf{q}_{l} ,
\end{equation} 
\begin{equation}
    \hat{\mathbf{g}}_{lb}^{\left[rj\right]} = \mathbf{R}
    _{lb}^{\left[rj \right]} \left( \mathbf{I}_{M} + \sum\limits_{u=1}^{C}\sum\limits_{v \in \nu \left( u \right)} \mathbf{R}_{lu}^{\left[ rv \right]} \right)^{-1}\mathbf{Y}_p^{\left[ r \right]} \mathbf{q}_{l} .
\end{equation}
Then,
\begin{align}
    &A  = \mathop{\mathbb{E}} \biggl[ \biggl | \hat{\mathbf{g}}_{lr}^{[rk]H} \mathbf{\Phi}_k \mathbf{\Phi}_k^H \hat{\mathbf{g}}_{lb}^{[rj]} \biggl |^{2}\biggl] \\
    & =  \mathop{\mathbb{E}} \biggl[ \biggl |\mathbf{q}_{l}^H\mathbf{Y}_{p}^{[r]H}\biggl(\mathbf{I}_{M} + \sum\limits_{u=1}^{C}\sum\limits_{v \in \nu(u)}\mathbf{R}_{lu}^{[rv]} \biggl)^{-1} \nonumber \\ & \times \mathbf{R}_{lr}^{[ rk]}  \mathbf{\Phi}_{k} \mathbf{\Phi}_{k}^{H} \mathbf{R}_{lb}^{[rj]}\biggl( \mathbf{I}_{M} + \sum\limits_{u=1}^{C}\sum\limits_{v \in \nu(u)}\mathbf{R}_{lu}^{[rv]} \biggl)^{-1} \mathbf{Y}_{p}^{[ r]} \mathbf{q}_{l} \biggl |^{2}\biggl] \\
    & =  Tr \Biggl\{ \left( \mathbf{I}_{M} + \sum\limits_{u=1}^{C}\sum\limits_{v \in \nu \left( u \right)} \mathbf{R}_{lu}^{\left[ rv \right]} \right)^{-1} \mathbf{R}_{lr}^{\left[ rk \right]} \mathbf{\Phi}_{k} \mathbf{\Phi}_{k}^{H}\mathbf{R}_{lb}^{\left[ rj \right]} \nonumber \\& \times \left( \mathbf{I}_{M} + \sum\limits_{u=1}^{C}\sum\limits_{v \in \nu \left( u \right)} \mathbf{R}_{lu}^{\left[ rv \right]} \right)^{-1}\mathbf{R}_{lb}^{\left[ rj \right]}\mathbf{\Phi}_{k} \mathbf{\Phi}_{k}^{H}\mathbf{R}_{lr}^{\left[ rk \right]} \Biggl\} \nonumber\\
    & +  \left | Tr \left\{ \left( \mathbf{I}_{M} + \sum\limits_{u=1}^{C}\sum\limits_{v \in \nu \left( u \right)} \mathbf{R}_{lu}^{\left[ rv \right]} \right)^{-1}\mathbf{R}_{lr}^{\left[ rk \right]}\mathbf{\Phi}_{k} \mathbf{\Phi}_{k}^{H}\mathbf{R}_{lb}^{\left[ rj \right]} \right\} \right| ^{2} \\
    & =  Tr \left\{ \mathbf{C}_{lr}^{\left[ rk \right]} \mathbf{\Phi}_{k} \mathbf{\Phi}_{k}^{H} \mathbf{C}_{lb}^{\left[ rj \right]}\mathbf{\Phi}_{k} \mathbf{\Phi}_{k}^{H} \right\} \nonumber \\ &\quad+ \left| Tr \left\{ \mathbf{\Theta}_{lb}^{\left[ rj \right]} \mathbf{R}_{lr}^{\left[ rk \right] }\mathbf{\Phi}_{k} \mathbf{\Phi}_{k}^{H}\right\} \right| ^{2}.
 \end{align}
 where the last two equalities are due to Lemma 3. 
 \newline
Finally, we have:
\begin{equation}
    \begin{aligned}
        \mathop{\mathbb{V}ar} \left(I_2\right) &\leq \sum\limits_{\substack{b=1 \\ b \neq r}}^C\sum\limits_{ j \in \nu \left( b \right)} \Bigg(Tr \left\{ \mathbf{C}_{lr}^{\left[ rk \right]} \mathbf{\Phi}_{k} \mathbf{\Phi}_{k}^{H} \mathbf{C}_{lb}^{\left[ rj \right]}\mathbf{\Phi}_{k} \mathbf{\Phi}_{k}^{H} \right\} \\&\qquad + \left| Tr \left\{ \mathbf{\Theta}_{lb}^{\left[ rj \right]} \mathbf{R}_{lr}^{\left[ rk \right] }\mathbf{\Phi}_{k} \mathbf{\Phi}_{k}^{H}\right\} \right| ^{2}\Bigg) \\
        &\qquad + \sum\limits_{\substack{ j \in \nu \left( r\right) \\ j \neq k}}\Bigg( Tr \left\{ \mathbf{C}_{lr}^{\left[ rk \right]} \mathbf{\Phi}_{k} \mathbf{\Phi}_{k}^{H} \mathbf{C}_{lr}^{\left[ rj \right]}\mathbf{\Phi}_{k} \mathbf{\Phi}_{k}^{H} \right\} \\&\qquad+ \left| Tr \left\{ \mathbf{\Theta}_{lr}^{\left[ rj \right]} \mathbf{R}_{lr}^{\left[ rk \right] }\mathbf{\Phi}_{k} \mathbf{\Phi}_{k}^{H}\right\} \right| ^{2} \Bigg) \alpha^4 .
    \end{aligned}
\end{equation}

Let us now consider the term $I_3$, which is caused by other users assigned orthogonal pilot sequences. While this component of interference vanishes asymptotically with a large number of BS antennas, it contributes to the total interference in the finite-antenna regime. This is due to the residual non-orthogonality between the channel estimates $\hat{\mathbf{g}}_{lr}^{[rk]}$ and $\hat{\mathbf{g}}_{ib}^{[rj]}$.
\begin{align}
        \mathop{\mathbb{V}ar}\left( I_3\right) &= \mathop{\mathbb{V}ar} \left( \sum\limits_{b=1}^{C}\sum\limits_{j \in \nu \left( b \right)}\sum\limits_{\substack{i=1 \\ i \neq l}}^
        {\tau}  \hat{\mathbf{g}}_{lr}^{\left[ rk \right]H} \mathbf{\Phi}_k \mathbf{\Phi}_k^H \hat{\mathbf{g}}_{ib}^{\left[ rj \right]} s_{jib} \right) \\
        &= \sum\limits_{b=1}^{C} \sum\limits_{j \in \nu \left( b \right)} \sum\limits_{\substack{i=1 \\ i \neq l}}^{\tau} \Bigg(\mathop{\mathbb{E}} \left[ \left | s_{jib} \right | ^2 \right] \nonumber \\ &\quad \times\mathop{\mathbb{E}} \left[ \hat{\mathbf{g}}_{lr}^{\left[ rk \right]H} \mathbf{\Phi}_k \mathbf{\Phi}_k^H \hat{\mathbf{g}}_{ib}^{\left[ rj \right]}\hat{\mathbf{g}}_{ib}^{\left[ rj \right]H}\mathbf{\Phi}_k \mathbf{\Phi}_k^H\hat{\mathbf{g}}_{lr}^{\left[ rk \right]}\right] \Bigg)\\
        &=\sum\limits_{b=1}^{C} \sum\limits_{j \in \nu \left( b \right)} \sum\limits_{\substack{i=1 \\ i \neq l}}^{\tau}Tr \Bigg\{ \mathop{\mathbb{E}} \left[ \hat{\mathbf{g}}_{lr}^{\left[ rk \right]}\hat{\mathbf{g}}_{lr}^{\left[ rk \right]H}\mathbf{\Phi}_k \mathbf{\Phi}_k^H \right] \nonumber \\ &\qquad \times \mathop{\mathbb{E}} \left[ \hat{\mathbf{g}}_{ib}^{\left[ rj \right]}\hat{\mathbf{g}}_{ib}^{\left[ rj \right]H}\mathbf{\Phi}_k \mathbf{\Phi}_k^H \right] \Bigg\} \\
        &=\sum\limits_{b=1}^{C} \sum\limits_{j \in \nu \left( b \right)} \sum\limits_{\substack{i=1 \\ i \neq l}}^{\tau}Tr \left\{ \mathbf{C}_{lr}^{\left[ rk \right]}\mathbf{\Phi}_k \mathbf{\Phi}_k^H  \mathbf{C}_{ib}^{\left[ rj \right]} \mathbf{\Phi}_k \mathbf{\Phi}_k^H \right\}.
\end{align}

The term $I_4$ is caused by the \textbf{estimation error} of the channel vectors. Since $\hat{\mathbf{g}}_{lr}^{[rk]}$ and $\tilde{\mathbf{g}}_{ib}^{[rj]}$ are statistically independent, we obtain:
\begin{align}
        \mathop{\mathbb{V}ar}\left( I_4 \right) &= \mathop{\mathbb{V}ar}\left(\sum\limits_{b=1}^{C}\sum\limits_{j \in \nu \left( b \right)}\sum\limits_{i=1}^{\tau}\hat{\mathbf{g}}_{lr}^{\left[ rk\right] H} \mathbf{\Phi}_k \mathbf{\Phi}_k^H\tilde{\mathbf{g}}_{ib}^{\left[rj\right]} s_{jib} \right) \\
        &=\sum\limits_{b=1}^{C}\sum\limits_{j \in \nu \left( b \right)}\sum\limits_{i=1}^{\tau}\mathop{\mathbb{E}}\left[ \left| s_{jib}\right|^2\right] \nonumber\\
        & \qquad\times\mathop{\mathbb{E}}\left[ \hat{\mathbf{g}}_{lr}^{\left[ rk \right] H}\mathbf{\Phi}_k \mathbf{\Phi}_k^H\tilde{\mathbf{g}}_{ib}^{\left[rj\right]} \tilde{\mathbf{g}}_{ib}^{\left[rj\right]H}\mathbf{\Phi}_k \mathbf{\Phi}_k^H \hat{\mathbf{g}}_{lr}^{\left[ rk \right] }\right] \\
        &= \sum\limits_{b=1}^{C}\sum\limits_{j \in \nu \left( b \right)}\sum\limits_{i=1}^{\tau} Tr\Bigg\{ \mathop{\mathbb{E}}\left[ \hat{\mathbf{g}}_{lr}^{\left[ rk \right]} \hat{\mathbf{g}}_{lr}^{\left[ rk \right] H}\mathbf{\Phi}_k \mathbf{\Phi}_k^H \right] \nonumber\\ &\qquad\times\mathop{\mathbb{E}}\left[ \tilde{\mathbf{g}}_{ib}^{\left[ rj \right] } \tilde{\mathbf{g}}_{ib}^{\left[ rj \right] H}\mathbf{\Phi}_k \mathbf{\Phi}_k^H \right] \Bigg\} \\
        &=\sum\limits_{b=1}^{C}\sum\limits_{j \in \nu \left( b \right)}\sum\limits_{i=1}^{\tau} Tr \Bigg\{ \mathbf{C}_{lr}^{\left[ rk \right] }\mathbf{\Phi}_k \mathbf{\Phi}_k^H \nonumber\\ &\qquad\qquad\times\left( \mathbf{R}_{ib}^{\left[ rj\right]} - \mathbf{C}_{ib}^{\left[ rj \right]} \right)\mathbf{\Phi}_k \mathbf{\Phi}_k^H \Bigg\}.
    \end{align}

Finally, the term $I_5$ accounts for the \textbf{additive noise}. Since the noise vector $\mathbf{n}_u \sim \mathcal{CN}(\mathbf{0}, \sigma^2\mathbf{I}_M)$ and the channel estimate $\hat{\mathbf{g}}_{lr}^{\left[rk \right]H}$ are statistically independent, $\mathop{\mathbb{V}ar}\left(I_5\right)$ is given by:
\begin{align}
        \mathop{\mathbb{V}ar}\left(I_5\right) &= \mathop{\mathbb{V}ar}\left( \hat{\mathbf{g}}_{lr}^{\left[rk \right]H} \mathbf{\Phi}_k \mathbf{\Phi}_k^H \mathbf{n}_u \right) \\
        &= \sigma^2 Tr \left\{ \mathop{\mathbb{E}} \left[ \hat{\mathbf{g}}_{lr}^{\left[rk \right]}\hat{\mathbf{g}}_{lr}^{\left[rk \right]H}\mathbf{\Phi}_k \mathbf{\Phi}_k^H \right] \right\} \\
        &=\sigma^2 Tr \left\{ \mathbf{C}_{lr}^{\left[ rk \right]} \mathbf{\Phi}_k \mathbf{\Phi}_k^H \right\}.
\end{align}
Therefore, with the matched filter receiver $\left( \mathbf{\Phi}_k \mathbf{\Phi}_k^H\hat{\mathbf{g}}_{lr}^{\left[ rk \right]} \right)^H$, the uplink SINR in the finite-antenna regime is given by \eqref{SINRexpr}.

\begin{lem}\label{lemma:trace_identity}
Consider the vector $\mathbf{a} \sim\ \mathcal{CN}(\mathbf{0},\ \mathbf{A})$, with covariance matrix $\mathbf{A}\in\ \mathbb{C}^{N\times N}$, and any diagonalizable matrix $\mathbf{B}\in\ \mathbb{C}^{N\times N}$. It holds that:
\begin{equation}
    \mathbb{E} \left\{ \left| \mathbf{a}^H \mathbf{Ba} \right| ^2 \right\} = \left| Tr \left( \mathbf{BA} \right) \right| ^2 + Tr \left( \mathbf{BA} \mathbf{B}^H \mathbf{A} \right) .
\end{equation}
\end{lem}
\begin{proof}
Let $\mathbf{w}=\mathbf{A}^{-\frac{1}{2}}\mathbf{a}$, which implies $\mathbf{a}=\mathbf{A}^{\frac{1}{2}}\mathbf{w}$ where $\mathbf{w} \sim\ \mathcal{CN}(\mathbf{0},\ \mathbf{I}_N)$. Thus, we can write: 
\begin{equation}
\mathbb{E}\left\{\left|\mathbf{a}^H\mathbf{Ba}\right|^2\right\}=\mathbb{E}\left\{\left|\mathbf{w}^H\left(\mathbf{A}^H\right)^\frac{1}{2}\mathbf{B}\mathbf{A}^\frac{1}{2}\mathbf{w}\right|^2\right\} .
\end{equation}
Next, let $\mathbf{U\Lambda}\mathbf{U}^H=\left(\mathbf{A}^H\right)^{\frac{1}{2}}\mathbf{B}\mathbf{A}^{\frac{1}{2}}$ denote the eigenvalue decomposition with $\mathbf{\Lambda}=diag\left(\lambda_1,\ldots,\lambda_N\right)$, and define $\widetilde{\mathbf{w}}=\left[{\widetilde{w}}_1,\ldots,{\widetilde{w}}_N\right]^T=\mathbf{U}^H\mathbf{w}\ \sim\ \mathcal{CN}(\mathbf{0},\ \mathbf{I}_N)$. It then follows that:
\begin{equation}
\begin{aligned}
\mathbb{E}&\left\{\left|\mathbf{w}^H\left(\mathbf{A}^H\right)^\frac{1}{2}\mathbf{B}\mathbf{A}^\frac{1}{2}\mathbf{w}\right|^2\right\} \nonumber\\ &=\mathbb{E}\left\{\left|{\widetilde{\mathbf{w}}}^H\ \mathbf{\Lambda}\widetilde{\mathbf{w}}\right|^2\right\} \\
        &=  \mathbb{E}\left\{\left|\sum_{n=1}^{N}{\left|{\widetilde{w}}_n\right|^2\lambda_n}\right|^2\right\} \\
        &=\sum_{n_1=1}^{N}\sum_{n_2=1}^{N}{\ \mathbb{E}\left\{\left\{\left|{\widetilde{w}}_{n_1}\right|^2\left|{\widetilde{w}}_{n_2}\right|^2\right\}\lambda_{n_1}\lambda_{n_2}^\ast\right\}} \\
        &\underset{=}{(a)} \sum_{n_1=1}^{N}\sum\limits_{\substack{n_2=1 \\ n_2 \neq n_1}}^{N} \lambda_{n_1}\lambda_{n_2}^\ast + \sum\limits_{n_1=1}^N 2\left|\lambda_{n_1} \right|^2 \\
        &= \sum_{n_1=1}^{N}{\lambda_{n_1}\sum_{n_2=1}^{N}\lambda_{n_2}^\ast}+\sum_{n_1=1}^{N}\left|\lambda_{n_1}\right|^2 \\
        &\underset{=}{(b)}  \left|Tr(\mathbf{\Lambda})\right|^2+Tr\left(\mathbf{\Lambda}\mathbf{\Lambda}^H\right) \\
        &= \left|Tr(\left(\mathbf{A}^H\right)^\frac{1}{2}\mathbf{B}\mathbf{A}^\frac{1}{2})\right|^2\nonumber\\&\qquad+Tr\left(\left(\mathbf{A}^H\right)^\frac{1}{2}\mathbf{B}\mathbf{A}^\frac{1}{2}\left(\mathbf{A}^H\right)^\frac{1}{2}\mathbf{B}\mathbf{A}^\frac{1}{2}\right) \\
        &\underset{=}{(c)} \left|Tr(\mathbf{BA})\right|^2+\ Tr\left(\mathbf{B}{\mathbf{AB}}^H\mathbf{A}\right) .
     \end{aligned}
 \end{equation}  
Here, $(a)$ follows from the independence between the elements of $\widetilde{\mathbf{w}}$ and the moments $\mathbb{E}[|\widetilde{w}_n|^2]=1$ and $\mathbb{E}[|\widetilde{w}_n|^4]=2$. In $(b)$, we rewrite the sum of the eigenvalues as the trace and reintroduce the unitary matrices from the eigenvalue decomposition. Finally, $(c)$ follows from the cyclic property of the trace, $\text{Tr}(\mathbf{C}_1\mathbf{C}_2) = \text{Tr}(\mathbf{C}_2\mathbf{C}_1)$, for any conformable matrices $\mathbf{C}_1$ and $\mathbf{C}_2$.
\end{proof}

\section{Proof of Theorem \ref{thm_nphard}}\label{AP:oproofNPhard}

We prove the NP-hardness of the resource allocation problem by demonstrating that it remains NP-hard even under the following restricted case, denoted as $(P)$:
\begin{enumerate}
    \item The network comprises exactly three cells ($C=3$).
    \item Only one user cluster is formed per cell.
    \item The SINR expression is given by Lemma \ref{lemme:sinr}.
    \item A single modulation scheme and a single pilot sequence are used by all users.
    \item All users have a strict minimum rate requirement, limiting each user to exactly one allocated frequency.
    \item Users within the same cell cannot share the same frequency. This condition is enforced by setting $\theta > 1$ and assuming that for all cells $b \in \{1,2,3\}$:
    \begin{equation}
        \begin{cases} 
            \beta_{1b}^{[b]} > \beta_{2b}^{[b]} > \ldots > \beta_{nb}^{[b]} \\ 
            \frac{\beta_{ib}^{[b]}}{\beta_{nb}^{[b]}} < \theta, \quad \forall i \in \{1, \ldots, n-1\} .
        \end{cases}
    \end{equation}
\end{enumerate}
Consequently, problem $(P)$ reduces to the following frequency allocation problem $(P')$:
\begin{equation}
\begin{aligned}
 \max_{\textbf{x}}& \sum\limits_{b=1}^{3}\sum\limits_{i=1}^{n}{w_{ib}\sum\limits_{f=f_1}^{f_n}{x_{ib}^{f}}} \\
\textrm{s.t.}  &\sum\limits_{i=1}^{n}x_{ib}^{f}  \leq  1, \; \forall b = \left\{ 1, 2, 3 \right\}; \;  \forall f = \left\{ f_1, \ldots, f_n \right\},  \\
&{SINR}_{ib}^{f}  \geq  \theta x_{ib}^{f}, \; \forall b = \left\{ 1, 2, 3 \right\}; \; \forall i = \left\{ 1, \ldots, n \right\}; \\ & \qquad\qquad\qquad\qquad \forall f = \left\{ f_1, \ldots fn \right\},
\end{aligned}
\end{equation}
where $x_{ib}^{f} \in \{0,1\}$ is a binary variable indicating whether user $i$ in cell $b$ is allocated frequency $f$. Applying the fourth condition to Lemma \ref{lemme:sinr}, the SINR expression in the second constraint simplifies to:
\begin{equation}
SINR_{ib}^{f}=\frac{\left( \beta_{ib}^{\left[ b \right]}\right)^{2}}{\sum\limits_{\substack{b^\prime=1 \\ b^\prime \neq b}}^3 \sum\limits_{j=1}^{n}{x_{j{b^\prime}}^{f} \left( \beta_{j{b^\prime}}^{\left[ b \right]} \right)^2}} .
\end{equation}

We perform a reduction from the 3-Dimensional Matching (3DM) problem, a well-known NP-complete problem \cite{garey}.

\textbf{3DM Problem with size n:} Consider three disjoint sets $X$, $Y$, and $Z$ such that $|X| = |Y| = |Z| = n$, and a subset $R $ of $ X \times Y \times Z$. The 3DM problem asks whether there exists a matching $\mathcal{M} \subseteq R$ satisfying:
\begin{enumerate}
    \item[(a)] For any two distinct triplets $(x_1, y_1, z_1), (x_2, y_2, z_2) \in \mathcal{M}$, we have $x_1 \neq x_2$, $y_1 \neq y_2$, and $z_1 \neq z_2$.
    \item[(b)] $X = \{x \mid (x,y,z) \in \mathcal{M}\}$, $Y = \{y \mid (x,y,z) \in \mathcal{M}\}$, and $Z = \{z \mid (x,y,z) \in \mathcal{M}\}$.
\end{enumerate}
Let us consider an arbitrary instance of the 3DM problem. Using the same notation, for all $\left(x_i,y_j,z_k\right)\in R$, where $x_i\in X$, $y_j\in Y$, $z_k\in Z$, and $i,j,k \in \{1,2,\ldots,n\}$, we define:
\begin{align*}
\beta_{i1}^{\left[1\right]}&=\beta_{j2}^{\left[2\right]}=\beta_{k3}^{\left[3\right]}=1,\\ \beta_{j2}^{\left[1\right]}&=\beta_{k3}^{\left[1\right]}=\beta_{i1}^{\left[2\right]}=\beta_{k3}^{\left[2\right]}=\beta_{i1}^{\left[3\right]}=\beta_{j2}^{\left[3\right]}= \frac{1}{2} .
\end{align*}
For all $\left(x_i,y_j,z_k\right)\notin R$, where $x_i\in X$, $y_j\in Y$, $z_k\in Z$, and $i,j,k \in \left\{1,2,\ldots,n\right\}$, we define:
\begin{align*}
\beta_{i1}^{\left[1\right]}&=\beta_{j2}^{\left[2\right]}=\beta_{k3}^{\left[3\right]}=1,\\
\beta_{j2}^{\left[1\right]}&=\beta_{k3}^{\left[1\right]}=\beta_{i1}^{\left[2\right]}=\beta_{k3}^{\left[2\right]}=\beta_{i1}^{\left[3\right]}=\beta_{j2}^{\left[3\right]}= \frac{1}{\sqrt{2}}.
\end{align*}
The corresponding decision instance $ \left( P_d \right)$ of problem $\left(P'\right)$ is given by:
\begin{equation}
\begin{aligned}
\left\{\begin{matrix} \sum\limits_{b=1}^{3}\sum\limits_{i=1}^{n}\sum\limits_{f=f_1}^{f_n}{x_{ib}^{f}} &\geq& \Omega , \\ \sum\limits_{i=1}^{n}{x_{ib}^{f}}  &\leq& 1, \; \forall b,f ,\\ 2{x_{ib}^{f}} \left( \sum\limits_{\substack{b^\prime=1 \\ b^\prime \neq b}}^{3} \sum\limits_{j=1}^{n} { x_{j{b^\prime}}^{f} \left( \beta_{j{b^\prime}}^{\left[ b \right]} \right)^2 }  \right) &\leq& \left( \beta_{ib}^{\left[ b \right]} \right)^{2}&, \; &\forall i, b, f,         \\\end{matrix}\right. 
\end{aligned}
\end{equation}
where we set the following parameters:
\begin{equation*}
w_{ib} = 1,\; \forall i=\left\{ 1, 2, \ldots, n\right\}; \forall b=\left\{1,2,3\right\} ,
\end{equation*}
\begin{equation*}
\theta=2,
\end{equation*}
\begin{equation*}
\Omega=3n.
\end{equation*}
Now, we will show that the two instances are equivalent (i.e., the answer to the 3-Dimensional matching problem is yes if and only if the answer to the constructed decision problem $(P_d)$ is yes). First, we show that if the answer to the 3-Dimensional matching problem is yes, then the answer to problem $(P_d)$ is yes.

Consider the following matching $\mathcal{M}\subseteq R$ such that $\mathcal{M}=\left(t_l\right)_{l=1}^n$, where the $t_l$ are triplets of elements belonging to $X$, $Y$, and $Z$ respectively. Now, let us consider the following $\left(P_d\right)$ solution:
\begin{itemize}
\item $\forall\ l$, the triplet $t_l$ corresponds to three users from $X$, $Y$, and $Z$ that are using the same frequency $f_l$.
\item $\forall\ l$, let us consider $i,j,k\ \in\left\{1,2,\ldots,n\right\}$ such that $x_i\in X$, $y_j\in Y$, and $z_k\in Z$ are in some triplet $t_l=\left(x_i,y_j,z_k\right)$.
\end{itemize}

Now, we set $x_{i1}^{f_l}=1,\ x_{j2}^{f_l}=1,\ x_{k3}^{f_l}=1$. Given that $\left(x_i,y_j,z_k\right) \in R$, the following inequalities are verified:
\begin{equation*}
2{x_{i1}^{f_l}} \left( x_{j2}^{f_l} \left( \beta_{j2}^{\left[ 1 \right]} \right)^2 + x_{k3}^{f_l} \left( \beta_{k3}^{\left[ 1 \right]} \right)^2 \right) \leq \left( \beta_{i1}^{\left[ 1 \right]} \right)^2,
\end{equation*}
\begin{equation*}
2{x_{j2}^{f_l}} \left( x_{i1}^{f_l} \left( \beta_{i1}^{\left[ 2 \right]} \right)^2 + x_{k3}^{f_l} \left( \beta_{k3}^{\left[ 2 \right]} \right)^2 \right) \leq \left( \beta_{j2}^{\left[ 2 \right]} \right)^2,
\end{equation*}
\begin{equation*}
2{x_{k3}^{f_l}} \left( x_{i1}^{f_l} \left( \beta_{i1}^{\left[ 3 \right]} \right)^2 + x_{j2}^{f_l} \left( \beta_{j2}^{\left[ 3 \right]} \right)^2 \right) \leq \left( \beta_{k3}^{\left[ 3 \right]} \right)^2.
\end{equation*}

Following this procedure, we obtain $3n$ variables with a value of one. The remaining $3n^2-3n$ Boolean variables are set to zero; that is, we set:
\begin{equation*}
\forall\ i^\prime=1,\ldots,n;\ i^\prime\neq i \text{ we have } x_{i^\prime 1}^{f_l}=0,
\end{equation*}
\begin{equation*}
\forall\ j^\prime=1,\ldots,n;\ j^\prime\neq j \text{ we have } x_{j^\prime 2}^{f_l}=0,
\end{equation*}
\begin{equation*}
\forall\ k^\prime=1,\ldots,n;\ k^\prime\neq k \text{ we have } x_{k^\prime 3}^{f_l}=0,
\end{equation*}
and the following $3\left(n-1\right)$ inequalities are satisfied:
\begin{equation*}
\forall\ i^\prime=1,\ldots,n;\ i^\prime\neq i; \; \left( \beta_{{i^\prime 1}}^{\left[1\right]} \right)^2 \geq 0,
\end{equation*}
\begin{equation*}
\forall\ j^\prime=1,\ldots,n;\ j^\prime\neq j; \; \left( \beta_{{j^\prime 2}}^{\left[2\right]} \right)^2 \geq 0,
\end{equation*}
\begin{equation*}
\forall\ k^\prime=1,\ldots,n;\ k^\prime\neq k; \; \left( \beta_{{k^\prime 3}}^{\left[k\right]} \right)^2 \geq 0.
\end{equation*}
Therefore, $\forall\ l=\left\{1,2,\ldots,n\right\}$ and $\ \forall\ b=\left\{1,2,3\right\}$, we have $\sum\limits_{i=1}^{n}x_{ib}^{f_l}=1$.
Thus, we have found a solution for problem $\left(P_d\right)$ that achieves $\sum\limits_{b=1}^{3}\sum\limits_{i=1}^{n}\sum\limits_{f=f_1}^{f_n}x_{ib}^f\geq3n$ with equality.

Now, we must show that if the answer to the decision problem $\left(P_d\right)$ is yes, then the answer to the 3DM problem is yes. Suppose that there exists a vector $\mathbf{x}$ of length $3n^2\times1$ that is a solution to the decision problem $\left(P_d\right)$. Let us construct the solution to the 3DM problem as follows:

\begin{itemize}
\item By the construction of the decision problem $\left(P_d\right)$, the first two conditions are:
\begin{equation*}
\sum\limits_{b=1}^{3}\sum\limits_{i=1}^{n}\sum\limits_{f=f_1}^{f_n}x_{ib}^f\geq3n,
\end{equation*}
\begin{equation*}
\sum\limits_{i=1}^{n}x_{ib}^f\le1;\ \forall\ b=\left\{1,2,3\right\},f=\left\{f_1,f_2,\ldots,f_n\right\} .
\end{equation*}
This implies that each user in the system has exactly one allocated frequency.
\item $\forall\ l \in \left\{1,2,\ldots,n\right\}$: $\exists\ i,j,k \in \left\{1,2,\ldots,n\right\}$ such that $x_{i1}^{f_l}=1,\ x_{j2}^{f_l}=1,\ x_{k3}^{f_l}=1$. Here, $i$, $j$, and $k$ correspond to three users from cells 1, 2, and 3 respectively that are using the same frequency $f_l$. We define the triplet $t_l=\left(x_i,y_j,z_k\right)$ such that $x_i=x_{i1}^{f_l}$, $y_j=x_{j2}^{f_l}$, and $z_k=x_{k3}^{f_l}$. Since $\left(x_{i1}^{f_l},\ x_{j2}^{f_l},x_{k3}^{f_l}\right)$ are users sharing the same frequency $f_l$, the following three inequalities of the SINR constraints are satisfied:
\begin{equation*}
2{x_{i1}^{f_l}} \left( x_{j2}^{f_l} \left( \beta_{j2}^{\left[ 1 \right]} \right)^2 + x_{k3}^{f_l} \left( \beta_{k3}^{\left[ 1 \right]} \right)^2 \right) \leq \left( \beta_{i1}^{\left[ 1 \right]} \right)^2,
\end{equation*}
\begin{equation*}
2{x_{j2}^{f_l}} \left( x_{i1}^{f_l} \left( \beta_{i1}^{\left[ 2 \right]} \right)^2 + x_{k3}^{f_l} \left( \beta_{k3}^{\left[ 2 \right]} \right)^2 \right) \leq \left( \beta_{j2}^{\left[ 2 \right]} \right)^2,
\end{equation*}
\begin{equation*}
2{x_{k3}^{f_l}} \left( x_{i1}^{f_l} \left( \beta_{i1}^{\left[ 3 \right]} \right)^2 + x_{j2}^{f_l} \left( \beta_{j2}^{\left[ 3 \right]} \right)^2 \right) \leq \left( \beta_{k3}^{\left[ 3 \right]} \right)^2.
\end{equation*}
The values 
\begin{equation*}
\begin{aligned}
\left\{\begin{matrix}  \beta_{i1}^{\left[1\right]}=\beta_{j2}^{\left[2\right]}=\beta_{k3}^{\left[3\right]}=1 ,\\ \beta_{j2}^{\left[1\right]}=\beta_{k3}^{\left[1\right]}=\beta_{i1}^{\left[2\right]}=\beta_{k3}^{\left[2\right]}=\beta_{i1}^{\left[3\right]}=\beta_{j2}^{\left[3\right]}= \frac{1}{\sqrt{2}}  ,\\\end{matrix}\right. 
\end{aligned}
\end{equation*} 
do not satisfy these three inequalities since $1\leq 2\left(\frac{1}{2}+\frac{1}{2}\right)$. Therefore, we must have
\begin{equation*}
\begin{aligned}
\left\{\begin{matrix}  \beta_{i1}^{\left[1\right]}=\beta_{j2}^{\left[2\right]}=\beta_{k3}^{\left[3\right]}=1 ,\\ \beta_{j2}^{\left[1\right]}=\beta_{k3}^{\left[1\right]}=\beta_{i1}^{\left[2\right]}=\beta_{k3}^{\left[2\right]}=\beta_{i1}^{\left[3\right]}=\beta_{j2}^{\left[3\right]}= \frac{1}{2}  ,\\\end{matrix}\right. 
\end{aligned}
\end{equation*}
and the triplet $t_l=\left(x_i,y_j,z_k\right)\in R$.
\item After this step, the triplets $\left(t_l\right)_{l=1}^n$ are constructed. For all $l, l^\prime$ such that $l\neq l^\prime$, with $t_l=\left(x_i, y_j, z_k\right)$ and $t_{l^\prime}=\left(x_{i^\prime}, y_{j^\prime},z_{k^\prime}\right)$, we have $x_i\neq x_{i^\prime}$, $y_j\neq y_{j^\prime}$, and $z_k\neq z_{k^\prime}$. Therefore, $\mathcal{M}=\left(t_l\right)_{l=1}^n\subseteq R$ is a valid matching for the 3DM problem.
\end{itemize}
Thus, we have proven the equivalence between any instance of the 3DM problem and an instance of problem $\left(P^\prime\right)$. We can conclude that problem $\left(P^\prime\right)$ and the original optimization problem $\left( P \right)$ are NP-hard. Furthermore, since the chosen NP-complete problem is strongly NP-complete, the optimization problem is strongly NP-hard.

\section{Proof of Theorem \ref{theorem:polynomialcase}} \label{Ap:proofPolyn}

To prove that the problem is solvable in polynomial time, we show that it can be formulated as a LP in canonical form:
\begin{equation}
\begin{array}{rrclcl}
\displaystyle \max_{\textbf{x}} & \mathbf{c}^T\mathbf{x} \\
\textrm{s.t.} & \mathbf{A}\mathbf{x} & \leq & \mathbf{b} ,\label{mip:unimodulaire}
\end{array}
\end{equation}
where $\mathbf{A}$ is a totally unimodular matrix (i.e., the determinant of every square submatrix of $\mathbf{A}$ is $0$, $1$, or $-1$), and $\mathbf{b}$ is an integer vector. Under these conditions, the LP is guaranteed to yield an optimal integer solution and can be solved in polynomial time.

The first constraint of the optimization problem \eqref{mip:1modulation} is equivalent to (see Claim 1):
\begin{equation}
\begin{aligned}
x_{ib}^{f{m_1}}+ x_{ib}^{f{m_2}} \leq 1; \; &\forall b \in \left\{1,2\right\}, \; \forall i \in \left\{1,2, \ldots,n \right\}, \\ &\forall f \in \left\{1,2, \ldots,n \right\},\\& \forall m_1,m_2 \in \left\{1,2, \ldots,L\right\} \textnormal{and } m_1 \neq m_2 .
\end{aligned}
\end{equation}

The SINR constraint \eqref{mip:sinr} can be written as follows for a given $i$, $b$, $f$, and $m$:
\begin{equation}
\left( \beta_{ib}^{\left[ b \right]}\right)^2 \geq {\theta_m}{x_{ib}^{fm}}{\left( \sum\limits_{j=1}^{n}{\left( \beta_{j{b^\prime}}^{\left[ b \right]}\right)^2x_{j{b^\prime}}^{fm}}\right)} .
\end{equation}

These constraints can be equivalently formulated by considering pairs of variables. For user $i$ in cell $b$ and user $j$ in the interfering cell $b^\prime$ (where $b^\prime \neq b$):
\begin{enumerate}
    \item If $\left( \beta_{ib}^{\left[ b \right]}\right)^2 < \theta_m \left( \beta_{j{b^\prime}}^{\left[ b \right]}\right)^2$, then users $i$ and $j$ cannot be allocated the same frequency $f$ using modulation $m$. Thus, we can express this constraint as: $x_{ib}^{fm}+ x_{j{b^\prime}}^{fm} \leq 1$.
    \item If $\left( \beta_{ib}^{\left[ b \right]}\right)^2 \geq \theta_m \left( \beta_{j{b^\prime}}^{\left[ b \right]}\right)^2$, then there is no conflict, and therefore no additional constraint on the variables $x_{ib}^{fm}$ and $x_{j{b^\prime}}^{fm}$.
\end{enumerate}

Thus, the problem reduces to the canonical form \eqref{mip:unimodulaire} with a binary vector $\mathbf{b}$. Here, $\mathbf{A}$ serves as the incidence matrix of a conflict graph: vertices represent the variables $x_{ib}^{fm}$, and edges connect mutually exclusive variables (e.g., $x_{ib}^{f{m_1}}+ x_{ib}^{f{m_2}} \leq 1$). Crucially, this special case is characterized precisely by the fact that all SINR constraints reduce to pairwise exclusivity between users in different cells. Because there are only two cells ($C=2$) and intra-cell interference is zero, the vertices can be partitioned into two disjoint sets, making the conflict graph strictly bipartite. The incidence matrix of any bipartite graph is totally unimodular; therefore, solving the relaxed LP guarantees an optimal integer solution in polynomial time.

\begin{claim}
Consider a vector $\mathbf{x}=\left( x_1, x_2, \ldots, x_L\right)$ of length $L$ with entries 0 or 1. Then, the following holds:
\begin{equation}
\sum\limits_{i=1}^L {x_i} \leq 1 \iff x_i + x_j \leq 1; \; \forall i,j \in \left\{1,2, \ldots,L \right\} , i \neq j.
\end{equation}
\end{claim}
\begin{proof}
First, we prove that if $\sum\limits_{i=1}^L {x_i} \leq 1$, then we have $x_i + x_j \leq 1; \; \forall i,j/i \neq j$. Let $\mathbf{x}=\left( x_1, x_2, \ldots, x_L\right)$ be a vector such that $\sum\limits_{i=1}^L {x_i} \leq 1$, i.e., at most one component of this vector is 1. If such a component exists, let $m^\ast$ be its index. Then:
\begin{itemize}
\item $\forall i \in \left\{1,2, \ldots,L \right\}, \; i \neq m^\ast; \; x_i + x_{m^\ast} = 1.$
\item $\forall i,\,j \in \left\{1,2, \ldots,L \right\}, \; i \neq j, \; i \neq m^\ast \textnormal{and } j \neq m^\ast; \; x_i + x_j =0 .$
\end{itemize}
Thus, we have that $\forall i,j \in \left\{1,2, \ldots,L \right\} \textnormal{and } i \neq j; \; x_i + x_j \leq 1$. Next, we prove that if $\forall i,j \in \left\{1,2, \ldots,L \right\} \textnormal{and } i \neq j; \; x_i + x_j \leq 1$, then we have $\sum\limits_{i=1}^L x_i \leq 1$. Suppose that $\exists i^\ast, j^\ast \in \left\{1,2, \ldots,L \right\} \textnormal{and } i^\ast \neq j^\ast$ such that $x_{i^\ast}=1$, $x_{j^\ast}=1$. Then, $x_{i^\ast} + x_{j^\ast} > 1$. Thus, at most one component of the vector $\mathbf{x}$ can be 1, and we have $\sum\limits_{i=1}^L x_i \leq 1$.
\end{proof}

\bibliographystyle{IEEEtran}

\end{document}